\documentclass[11pt,a4paper]{article}
\usepackage[T1]{fontenc}
\usepackage[utf8]{inputenc}
\usepackage{authblk}
\usepackage{amsmath}
\usepackage{graphicx}
\usepackage{wrapfig}
\usepackage{amsmath}
\usepackage{times}
\usepackage{amssymb}
\usepackage{array}
\usepackage{latexsym}
\usepackage{amsthm}
\usepackage{amsfonts}
\usepackage{nomencl}
\makenomenclature
\usepackage{cite}
\usepackage{array,tabularx}
\usepackage{graphicx}
\usepackage{amsmath}
\usepackage{multirow}
\usepackage{float}
\usepackage{caption}
\usepackage{subcaption}
\newenvironment{conditions*}
  {\par\vspace{\abovedisplayskip}\noindent
   \tabularx{\columnwidth}{>{$}l<{$} @{${}={}$} >{\raggedright\arraybackslash}X}}
  {\endtabularx\par\vspace{\belowdisplayskip}}

\date{}
\mathchardef\mhyphen="2D
\newcommand{\bea}{\begin{equation}}
\newcommand{\eea}{\end{equation}}

\newcommand{\bee}{\begin{equation*}}
\newcommand{\eee}{\end{equation*}}

\numberwithin{equation}{section}
\newcounter{saveeqn}

\begin{document}
\title{
\textbf{Effect of surfactant concentration and interfacial slip on the flow past a viscous drop at low surface P\'{e}clet number}}
\author[1]{G.P. Raja Sekhar \thanks{rajas@maths.iitkgp.ernet.in}}
\author[1]{V. Sharanya \thanks{sharanya@iitkgp.ac.in}}
\author[2]{Christian Rohde}
\affil[1]{Department of Mathematics, Indian Institute of Technology,
              Kharagpur-721302, India}
\affil[2]{Institute of Applied Analysis and Numerical Simulation, University of Stuttgart}

\renewcommand\Authands{ and }
\maketitle
\date{\today}
\begin{abstract}
The motion of a viscous drop is investigated when the interface is
fully covered with a stagnant layer of surfactant in an arbitrary
unsteady Stokes flow for the  low surface P\'{e}clet number limit.
The effect of the interfacial slip coefficient on the behavior of
the flow field is also considered. The hydrodynamic problem is
solved by  the  solenoidal decomposition method and the drag force
is computed in terms of Faxen's laws using a perturbation ansatz
in powers of the surface  P\'{e}clet number. The analytical
expressions for the migration velocity of the drop are also
obtained in powers of the surface P\'{e}clet number. Further
instances corresponding to a given ambient flow as uniform flow,
Couette flow, Poiseuille flow are analyzed. Moreover, it is
observed that, a surfactant-induced cross-stream migration of the
drop occur towards the centre-line in both Couette flow and
Poiseuille flow cases. The variation of the drag force
and migration velocity is computed for  different parameters such
as P\'{e}clet number, Marangoni number etc.
\end{abstract}

\section{Introduction}
\label{intro}
The motion of drops and bubbles is a common phenomenon understanding which is important to realize many industrial and chemical
applications. Some properties such as deformability, inertia, and external (thermal or chemical) gradients influence the migration of drops.
The variation of
temperature or the presence of surfactants causes  variations in the interfacial gradient.
Young, Goldstein and Block \cite{young1959motion} were the first to study the flow past a drop by considering thermal
effects.
Subramanian and Balasubramaniam~\cite{subramanian2001motion} have computed the drag force in terms of Faxen's laws by considering
the thermal effects in an axisymmetric Stokes flow. Subramanian \cite{subramanian1983thermocapillary}
calculated the settling velocity of a drop by considering  thermal effects in a steady axisymmetric
flow. The unsteady motion of a vertically falling liquid drop in an axisymmetric flow has been
analyzed
by Chisnell \cite{chisnell1987unsteady}. Dill and Balasubramaniam \cite{dill1992unsteady} have studied the thermocapillary migration
of a drop in an axisymmetric unsteady Stokes flow. Choudhuri and Padmavathi \cite{choudhuri2014study} have calculated the drag and torque in
terms of Faxen's laws for an oscillatory Stokes flow past a drop. Choudhuri and Raja Sekhar \cite{choudhuri2013thermocapillary} have obtained the thermocapillary drift of a spherical
drop in a
steady arbitrary Stokes flow. Ramachandran et al. \cite{ramachandran2011effect}
discussed the impact of interfacial slip on the dynamics of a drop in a Stokes flow by using a
numerical approach based on the  boundary integral method. Ramachandran and Leal \cite{ramachandran2012effect} studied the
effect of interfacial slip on the drop deformation in a steady Stokes flow by using Navier slip boundary
conditions. Mandal et al. \cite{mandal2015effect} computed the shape of a drop by considering the interfacial
slip effect in an arbitrary steady Stokes flow by using Lamb's solution.

While these  works are  mostly on the migration of viscous drops in  pure ambient viscous flows, or in presence of thermocapillary effects,
there are also  studies concerned with the effect of surfactants on the motion of drops and
bubbles in  creeping flows. Surfactants are surface active agents that are adsorbed at a
fluid-fluid interface or at a liquid-gas interface, where they typically lower the interfacial
tension and cause a Marangoni effect.  It is observed that even a small amount of surfactant can
reduce the terminal velocity of a drop. For example, Levan and Newman \cite{leven1976effect} studied the effect of
surfactants on the terminal velocity of a drop in an axi symmetric flow.  Along the interface, the
 surfactant is governed by a convection-diffusion equation. Holbrook and LeVan \cite{holbrook1983retardationa}
 and Holbrook and LeVan \cite{holbrook1983retardationb} have used a collocation method to solve
the convection-diffusion problem for high P\'{e}clet numbers and studied the retardation
of drop motion when the surfactant is present. Sadhal and Johnson \cite{sadhal1983stokes} studied the flow
past a drop which is partially coated with a stagnant layer of surfactant for large surface
P\'{e}clet number. Many authors have examined the effect of soluble
and insoluble surfactants on the motion of drops using various numerical techniques
(Ref. \cite{oguz1988effects,alke2007vof}). Stone \cite{stone1990simple} derived a
convection-diffusion equation for the surfactant transport along a deforming interface.
Stone and Leal \cite{stone1990effects} used a numerical treatment to analyze the effect of surfactants on
the deformation and breakup of a drop. Hanna and Vlahovska \cite{hanna2010surfactant} discussed the
surfactant-induced migration of a drop in an unbounded Poiseuille flow for
large P\'{e}clet numbers.  A simplified CFD simulation was
performed to study the influence of surfactants on the rise of bubbles by
Fleckenstein and Bothe \cite{fleckenstein2013simplified}. Recently, Pak et al. \cite{pak2014viscous} calculated the migration
of a drop in a steady Poiseuille flow at low surface P\'{e}clet numbers.

The migration of a non-deforming clean spherical viscous drop at
zero Reynolds number in a pressure driven flow moves only along
the  flow direction (Ref. \cite{hetsroni1970flow}), i.e., there
can be no cross migration  in the absence of inertia and
deformation on a clean spherical drop. It is experimentally
observed that, for three dimensional Poiseuille flow and for
Couette flow, the migration due to deformation occurs towards the
center line (Ref.
\cite{goldsmith1962flow,karnis1967particle,chan1979motion}). The
cross migration due to inertial effects is also studied by many
authors (Ref. \cite{cox1968lateral,ho1974inertial}). It is also
found that the surfactant redistribution can also cause the cross
stream migration of drops (Ref.
\cite{hanna2010surfactant,stan2013magnitude,pak2014viscous}).
Recently, Mandal et al. \cite{mandal2015effect} have studied the
effect of interfacial slip on the cross migration of a drop in an
unbounded Poiseuille flow. However, these studies are restricted
to steady case and ambient Poiseuille flow. We are generalizing
the problem to an unsteady arbitrary ambient flow, by considering
the effects of interfacial slip as well as surfactant
concentration effects.

We are interested in the  case of arbitrary Stokes flow past drops  which
is  challenging due to its three  dimensional nature.
Note that the corresponding drag and torque can be obtained in a
compact form similar to Faxen's laws. For example, the recent study by
Choudhuri and Raja Sekhar \cite{choudhuri2013thermocapillary} discussed thermocapillary migration of a viscous spherical
drop and obtained the corresponding Faxen's laws. Consequently, Sharanya and Raja Sekhar \cite{sharanya2015thermocapillary}
have addressed thermocapillary migration of a spherical drop in an arbitrary unsteady Stokes flow.
We are motivated by these studies and consider the motion of a viscous spherical drop whose
interface is covered with a stagnant layer of surfactant in an arbitrary unsteady Stokes flow.
The arbitrary Stokes flow case is considered by Pak et al. \cite{pak2014viscous}, where they restrict the
flow to be steady, and the surfactant coating  the whole interface. The slip reduces the deformation of a drop in a
shear-type flow (Ref. \cite{ramachandran2011effect,ramachandran2012effect}). Also, it is noted that due to this slip condition the disturbance flow produced by a drop is expected to be
weakened in magnitude. In our present case,
we attempt a more generalized problem of an arbitrary transient Stokes flow past a drop for low
surface P\'{e}clet number. Also, we take into account the effect of interfacial
slip on the flow. We solve the problem for any given ambient flow and consider some special cases
to validate our results.

The objective of our present paper is to analyze  the behavior of the flow when the interfacial
slip effect and the surfactant concentration effect occurs for low surface P\'{e}clet numbers.
We use the solenoidal decomposition method to solve the
unsteady Stokes equations, which is motivated by
 the general solution  proposed by Venkatalaxmi et al. \cite{venkatalaxmi2004general}. We use slip boundary
conditions to see the effect of interfacial slip on the flow
behavior which has been  previously used by Ramachandran et al.
\cite{ramachandran2011effect} and Ramachandran and Leal
\cite{ramachandran2012effect}. If we denote the surfactant
concentration as $\Gamma$, we assume that  $\Gamma$ is governed by
a convection-diffusion equation
\cite{sadhal1983stokes,stone1990simple,wong1996surfactant}. We
find the surfactant concentration up to second order for an
arbitrary Stokes flow, i.e., up to O($Pe_s^2$) (Ref.
\cite{pak2014viscous}). We observe area-specific surfactant
distribution on the interface of the drop. We also solve for the
flow fields and obtain the settling velocity of the drop. We
compute migration velocity corresponding to surfactant coated drop
in Poiseuille flow and Couette flow and make some observations on
the cross flow migration.
\section{Problem Statement and Mathematical Formulation}\label{sec:model}
\begin{figure}
  \centerline{\includegraphics[width=12cm, height=10cm]{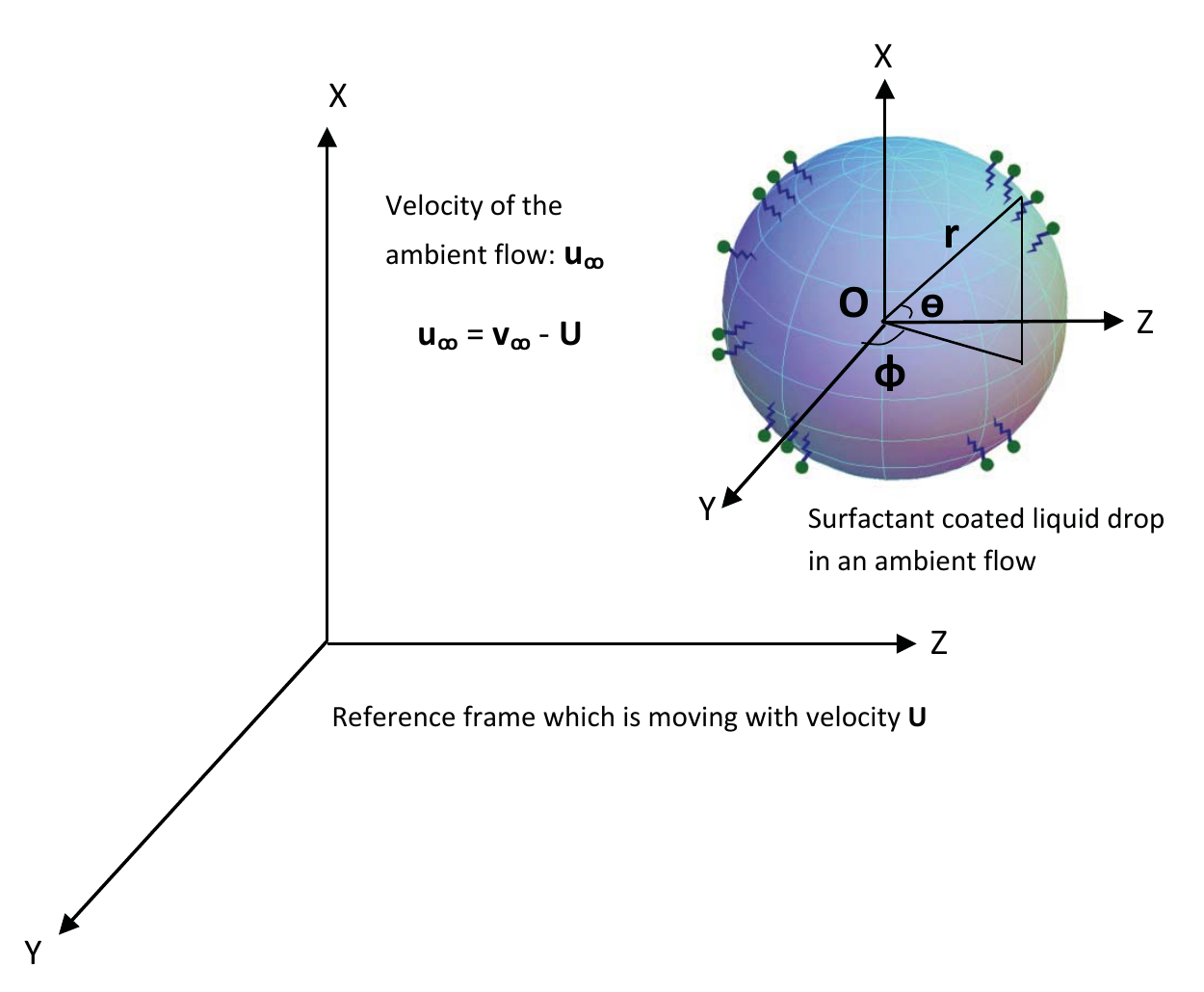}}
  \caption{Geometry of the problem}
\label{fig:geoprob}
\end{figure}

We consider the motion of a liquid drop of radius $a$ and viscosity $\mu^i$ in an unsteady Stokes flow, suspended in another unbounded Newtonian fluid of viscosity $\mu^e$ (see Fig. \ref{fig:geoprob}). Let the velocity of the fluid inside the drop be $\vec{v}^i$ and the velocity of the fluid outside the drop be $\vec{v}^e$. We assume that the settling velocity of the drop is ${\textbf{U}}$, which we determine later. The presence of a small amount of surface-active agents (surfactants) causes the variation in interfacial tension which influences the migration of the drop. We analyze the problem when the surfactant concentration effects and interfacial slip effects are considered. Surfactants are surface-active agents that lower the interfacial tension between two liquids. We neglect the inertial terms under negligible Reynolds number assumption. We assume a low surface P\'{e}clet number $Pe_s$. Further, we assume that the dimensional interfacial tension, $\sigma^*$, depends in an affine way on the dimensional surfactant concentration, $\Gamma^*$, i.e.,
\begin{equation}
\sigma^*=\sigma-R T \Gamma^*,
\nonumber
\end{equation}
where $\sigma$ is the interfacial tension when the interface is clean, $R$ is the gas constant and $T$ is the absolute temperature (Ref. \cite{pak2014viscous}).
We  non-dimensionalize the lengths by the drop radius $a$, velocities by the characteristic velocity scale of the background flow, $U_c$, time by its characteristic time scale $t_c$ and the surfactant concentration by its equilibrium value when the distribution is uniform, $\Gamma_{eq}$. The pressure is non dimensionalized by $\frac{\mu U_c}{a}$.

We assume that the flow inside and outside the drop is governed by the unsteady Stokes equations and the continuity equations which are given in the non dimensional form as follows:\\ \\
for $ r<1 $
\begin{eqnarray}
{\beta_i} \frac{\partial \vec{v}^i}{\partial t} = -\vec{\nabla} p^i + \vec{\nabla}^2 \vec{v}^i; & \mbox{ \,\,\,\, } \vec{\nabla} .\vec{v}^{i} = 0,
\label{eq:3}
\end{eqnarray}
and for $ r>1 $
\begin{eqnarray}
{\beta_e} \frac{\partial \vec{v}^e}{\partial t} = -\vec{\nabla} p^e + \vec{\nabla}^2 \vec{v}^e; & \mbox{ \,\,\,\,  } \vec{\nabla} .\vec{v}^{e} = 0.
\label{eq:4}
\end{eqnarray}
In the above
equations, ${\beta_e}=\frac{a^2}{{\nu_e} t_c}$ and ${\beta_i}=\frac{a^2}{{\nu_i} t_c}$ represent the
unsteadiness parameters corresponding to the flow inside and outside the drop respectively, which we assume to be unity, i.e.,
$t_c=\frac{a^2}{\nu_j}$.

We assume that the velocity field far from the drop approaches the undisturbed background flow, $\vec{v}^\infty$, i.e.,
\begin{equation}
\vec{v}^e \rightarrow \vec{v}_\infty\,\,\,\,\,\,\, as\,\,\,\,\,\, r \rightarrow \infty,
\label{eq:a1}
\end{equation}
which together with some pressure field $p_\infty$ satisfies the unsteady Stokes
and continuity equations.

The surfactant transport is governed by an unsteady convection-diffusion equation, (Ref. Stone \cite{stone1990simple} and Sadhal and Johnson \cite{sadhal1983stokes}), which is given in the non dimensional form as follows
\begin{eqnarray}
Pr_s\frac{\partial \Gamma}{\partial t}+Pe_s\left[\vec{\nabla}_s.(\Gamma \vec{v}_s)+\Gamma (\vec{v}.\hat{n})\vec{\nabla}_s.\hat{n}\right]=\vec{\nabla}_s^2 \Gamma,
\label{eq:diffusion}
\end{eqnarray}
where $\vec{v}_s=\vec{v^e}.\hat{t}$ is the velocity component tangential to the surface of the drop and $Pe_s=\frac{aU_c}{D_s}$ is the surface P\'{e}clet number which measures the importance of convection relative to diffusion. Here $D_s$ is the dimensional surface-diffusion constant. $Pr_s=\frac{\nu_e}{D_s}$ is the Prandtl number which is dimensionless and is defined as the ratio of momentum diffusivity to
surfactant diffusivity. Eq. (\ref{eq:diffusion}) includes the convective and diffusive contribution to the surfactant transport and a source-like contribution accounting for the variation of surfactant concentration resulting from the local changes in the interfacial area. (Ref. \cite{stone1990simple}).\\ \\
We solve the problem in a reference frame which is moving with the velocity of the drop, $\textbf{U}$, in which the drop appears to be stationary (see Fig. (\ref{fig:geoprob})). In this moving frame, the velocity fields inside and outside the drop are given respectively by
   \begin{eqnarray}
   \vec{u}^i=\vec{v}^i-\textbf{U},\nonumber
   \end{eqnarray}
    \begin{eqnarray}
   \vec{u}^e=\vec{v}^e-\textbf{U}.\nonumber
   \end{eqnarray}

One can observe that, these velocity fields also satisfy the unsteady Stokes and continuity equations given by\\ \\
for $ r<1 $
\begin{eqnarray}
\frac{\partial \vec{u}^i}{\partial t} = -\vec{\nabla} p^i + \vec{\nabla}^2 \vec{u}^i; & \mbox{ \,\,\,\, } \vec{\nabla} .\vec{u}^{i} = 0,
\label{eq:3a}
\end{eqnarray}
and for $ r>1 $
\begin{eqnarray}
\frac{\partial \vec{u}^e}{\partial t} = -\vec{\nabla} p^e + \vec{\nabla}^2 \vec{u}^e; & \mbox{ \,\,\,\,  } \vec{\nabla} .\vec{u}^{e} = 0.
\label{eq:4a}
\end{eqnarray}
The external velocity $\vec{u}^{e}$ is expected to meet the
following far field condition in the reference frame
\begin{equation}
\vec{u}^e \rightarrow \vec{u}_\infty=\vec{v}_\infty-\textbf{U}\,\,\,\,\,\,\, as\,\,\,\,\,\, r \rightarrow \infty.
\label{eq:a1}
\end{equation}
We follow the physical interpretations discussed by various authors~\cite{subramanian2001motion,happel1983low,cliftgrace,polanin2002hydrodynamics}  and adopt the following kinematic boundary conditions on the surface of the drop in non-dimensional form:\\ \\
{\textit{Vanishing normal component of the velocities, i.e.,}}
\begin{eqnarray}
\vec{u}^e.\hat{n}=0; & \mbox{$ ~~~~ $} \vec{u}^i.\hat{n} =0,
\end{eqnarray}
{\textit{Slip in the tangential component of velocities, i.e.,}}
\begin{eqnarray}
\vec{u}^e.\hat{t}-\vec{u}^i.\hat{t}=\alpha\tau^e_{\hat{n}\hat{t}},
\end{eqnarray}
{\textit{Tangential stress balance, i.e.,}}
\begin{eqnarray}
\tau^e_{\hat{n}\hat{t}}
-\mu\tau^i_{\hat{n}\hat{t}}=Ma\vec{\nabla}_s \Gamma.\hat{t},
\label{eq:jump.acap}
\end{eqnarray}
Since the stress fields and the surfactant concentration on the surface of the drop remain the same in both the laboratory frame and the moving frame, the tangential stress balance takes the same form as in both reference frames.
We note that the surfactant transport equation given in Eq. (\ref{eq:diffusion}) simplifies to
\begin{eqnarray}
Pr_s\frac{\partial \Gamma}{\partial t}+Pe_s\left[\vec{\nabla}_s.(\Gamma \vec{u}_s)\right]=\vec{\nabla}_s^2 \Gamma,
\label{eq:diffusiona}
\end{eqnarray}
in the moving reference frame. Here $\vec{u}_s$ is the velocity tangential to the surface of the drop in the moving frame.
\section{\label{methofsol}Method of solution}
We expand the velocity and pressure fields, surfactant concentration and migration velocity as a regular perturbation expansion for low surface P\'{e}clet number ($Pe_s \ll 1$), i.e.,
\begin{eqnarray}
\left[\vec{u}^e, \vec{u}^i, p^e, p^i, \Gamma, \textbf{U}\right]&=&\left[\vec{u}^e_0, \vec{u}^i_0, p^e_0, p^i_0, \Gamma_0, \textbf{U}_0\right]+Pe_s\left[\vec{u}^e_1, \vec{u}^i_1, p^e_1, p^i_1, \Gamma_1, \textbf{U}_1\right]\nonumber\\
&&+Pe_s^2\left[\vec{u}^e_2, \vec{u}^i_2, p^e_2, p^i_2, \Gamma_2, {\textbf{U}}_2\right]+O(Pe_s^3).
\label{eq:expansion}
\end{eqnarray}
Since the boundary value problem defined in Eqs. (\ref{eq:3a}) to (\ref{eq:jump.acap}) is independent of the perturbation parameter $Pe_s$, the velocity and pressure fields at all orders satisfy similar equations with the corresponding quantities as: leading order ($\vec{u}_0, p_0$), first order ($\vec{u}_1, p_1$), and second order ($\vec{u}_2, p_2$) etc. For brevity, we do not repeat these equations here.

\subsection{Representation of velocity}
By eliminating the pressure from the unsteady Stokes equations, one can verify that the velocity fields inside and outside the droplet satisfy
\begin{eqnarray}
\vec{\nabla}^2\left(\vec{\nabla}^2-\frac{\partial}{\partial
t}\right)\vec{u}^j=0\,\,\,\,\,\mbox{for\,\,\,} j=i,e. \label{eq:9}
\end{eqnarray}
By using the general solution
for the unsteady Stokes equation together with the equation of continuity, we can have the following representation for the velocity and pressure fields (see \cite{venkatalaxmi2004general})
\begin{eqnarray}
\vec{u}^j=\vec{\nabla} \times \vec{\nabla} \times (\textbf{r}\chi^j)+\vec{\nabla}\times(\textbf{r}\eta^j),
\label{eq:ref1}
\end{eqnarray}
\begin{eqnarray}
{p}^j=p^j_\infty+\rho_j\frac{\partial}{\partial
r}\left(\textbf{r}\left(\vec{\nabla}^2\chi^j-\frac{\partial
\chi^j}{\partial t}\right)\right),
\end{eqnarray}
where the scalars $\chi^j$ and $\eta^j$ are solutions of
\begin{eqnarray}
\vec{\nabla}^2\left(\vec{\nabla}^2-\frac{\partial}{\partial
t}\right)\chi^j=0, \label{eq:12}
\end{eqnarray}
\begin{eqnarray}
\left(\vec{\nabla}^2-\frac{\partial}{\partial
t}\right)\eta^j=0. \label{eq:13}
\end{eqnarray}
Here $\textbf{r}$ is the position vector and $p_\infty$ is a constant. Hence, the problem can now
 be handled in terms of the scalars $\chi^j$ and $\eta^j$.
 Accordingly, the boundary conditions in terms of $\chi^j$ and $\eta^j$ are given by\\ \\
{Vanishing normal component of the velocity}
\begin{eqnarray}
\chi^{e}=\chi^{i}=0 & \mbox{ on  $ r=1 $ }.
\label{eq:14}
\end{eqnarray}
{Slip in the tangential component of velocity}
\begin{eqnarray}
\frac{\partial \chi^{e}}{\partial r}-\frac{\partial \chi^{i}}{\partial r}=\alpha\frac{\partial^2 \chi^{e}}{\partial r^2},\,\,\,\,\,\, \eta^{e}-\eta^{i}=\alpha\frac{\partial}{\partial r}\left(\frac{\eta^e}{r}\right) & \mbox{ on  $ r=1 $ }.
\label{eq:15}
\end{eqnarray}
{Tangential stress balance}
\begin{eqnarray}
\frac{\partial}{\partial \theta}\left(\frac{\partial^2 \chi^{e}}{\partial r^2}-\mu\frac{\partial^2 \chi^{i}}{\partial r^2}\right)=Ma\frac{\partial \Gamma}{\partial \theta}& \mbox{ on  $ r=1 $ },
\label{eq:16a}
\end{eqnarray}
\begin{eqnarray}
\frac{\partial}{\partial \phi}\left(\frac{\partial^2 \chi^{e}}{\partial r^2}-\mu\frac{\partial^2 \chi^{i}}{\partial r^2}\right)=Ma\frac{\partial \Gamma}{\partial \phi}& \mbox{ on  $ r=1 $ },
\label{eq:16b}
\end{eqnarray}
\begin{eqnarray}
\frac{\partial}{\partial r}\left(\frac{\eta^e}{r}\right)=\mu\frac{\partial}{\partial r}\left(\frac{\eta^i}{r}\right) & \mbox{ on $r=1$}.
\label{eq:17}
\end{eqnarray}
{Finite velocity and pressure fields inside the drop require that}
\begin{eqnarray}
\chi^{i}<\infty,\,\,\,\,\,\,\, \eta^i<\infty.
\label{eq:18}
\end{eqnarray}
\subsection{Leading order problem}
The zeroth order surfactant transport equation corresponding to the general case given in (\ref{eq:diffusiona}) is
\begin{eqnarray}
Pr_s\frac{\partial \Gamma_0}{\partial t}=\vec{\nabla}_s^2 \Gamma_0\,.
\label{eq:diffusionb}
\end{eqnarray}
In order to obtain the leading order surfactant concentration $\Gamma_0$, we express $\Gamma_0$ in terms of spherical harmonics, i.e.,
\begin{eqnarray}
\Gamma_0=\sum\limits_{n=0}^\infty R_{n}^0(\theta,\phi)e^{-\lambda^2 t/Pr_s}\,,
\label{eq:difone}
\end{eqnarray}
where
\begin{eqnarray}
R_{n}(\theta,\phi)=\sum\limits_{m=0}^n
\left(E_{nm}^0\cos\,m\phi+F_{nm}^0\sin\,m\phi\right)P_n^m(\cos\,\theta),
\end{eqnarray}
are the spherical harmonics, $P_n^m(\eta)$ are associated Legendre polynomials and $E_{nm}^0$, $F_{nm}^0$ have to be determined such that $\Gamma_0$ satisfies (\ref{eq:diffusionb}).
Substituting the above expression (\ref{eq:difone}) in (\ref{eq:diffusionb}), we obtain $n(n+1)=-\lambda^2/Pr_s$. This is possible only when $n=0$ and $\lambda=0$ since we have $\lambda^2>0$. Therefore we have that $\Gamma_0$ is a constant,
which we take as unity, i.e., $\Gamma_0=1$.

%Further $\chi^e\rightarrow\chi_0$ and $\eta^e\rightarrow\eta_0$ as $r \rightarrow \infty$.
We represent the far-field ambient flow in terms of $\chi_0^\infty$ and $\eta_0^\infty$, given by
\begin{eqnarray}
\chi_0^\infty=\sum\limits_{n=1}^\infty \left[\alpha_{n}^0r^{n}+\beta_{n}^0f_n(\lambda_{e}r)\right]S_{n}(\theta,\phi)e^{\lambda_e^2 t},
\label{eq:19}
\end{eqnarray}
\begin{eqnarray}
\eta_0^\infty=\sum\limits_{n=1}^\infty \left[\gamma_{n}^0f_n(\lambda_{e}r)\right]T_{n}(\theta,\phi)e^{\lambda_e^2 t},
\label{eq:20}
\end{eqnarray}
where
\begin{eqnarray}
 S_n^0(\theta,\phi)=\sum\limits_{m=0}^n P_n^m(\eta)\left[A_{nm}^0\cos\, m\phi+B_{nm}^0\sin\, m\phi\right],
 \label{eq:h1}
 \end{eqnarray}
\begin{eqnarray}
 T_n^0(\theta,\phi)=\sum\limits_{m=0}^n P_n^m(\eta)\left[C_{nm}^0\cos\, m\phi+D_{nm}^0\sin\, m\phi\right],
 \label{eq:h2}
 \end{eqnarray}
 are spherical harmonics, and $\alpha_{n}^0$,
   $\beta_{n}^0$,  $\gamma_{n}^0$, $A_{nm}^0$, $B_{nm}^0$, $C_{nm}^0$ and $D_{nm}^0$ are the known
   coefficients. These coefficients are controlled by the choice of
   the ambient flow. For example, in case of uniform ambient flow, $\chi_0^\infty=\frac{1}{2}r \cos\,\theta e^{\lambda_e^2
   t},\,\,\eta_0^\infty=0$; and hence $\alpha_{1}^0=\frac{1}{2}$,
   $\alpha_{n}^0=0$ for $n\neq1$,
    $\beta_{n}^0=0$,  $\gamma_{n}^0=0$, $A_{10}^0=1$, $A_{nm}^0=0$ for $n\neq1$ or $m\neq0$, $B_{nm}^0=0$, $C_{nm}^0=0$ and
    $D_{nm}^0=0$.
 Here, $f_n(\lambda_j r)$ and $g_n(\lambda_j r)$ ($j=i, e$) are modified spherical Bessel function of first and second kind, respectively. Note that, for the bounded solution as $t\rightarrow\infty$, we require $\lambda_j^2<0$.
  In the presence of the spherical drop, the resultant flow due to the disturbance can be represented as general solution of
 Eqs.~(\ref{eq:12}) and (\ref{eq:13}) as follows,\\ \\
 for $r<1$
\begin{eqnarray}
\chi^i_0=\sum\limits_{n=1}^\infty \left[\bar{\alpha}_{n}^0r^{n}+\bar{\beta}_{n}^0f_n(\lambda_{i}r)\right]S_{n}^0(\theta,\phi)e^{\lambda_i^2 t},
\label{eq:21}
\end{eqnarray}
\begin{eqnarray}
\eta^i_0=\sum\limits_{n=1}^\infty \left[\bar{\gamma}_{n}^0f_n(\lambda_{i}r)\right]T_{n}^0(\theta,\phi)e^{\lambda_i^2 t},
\label{eq:22}
\end{eqnarray}
and for $r>1$
\begin{eqnarray}
%\begin{split}
\chi^e_0=\sum\limits_{n=1}^\infty \left[\alpha_{n}^0r^{n}+\frac{\hat{\alpha}_{n}^0}{r^{n+1}}+\beta_{n}^0f_n(\lambda_{e}r)+\hat{\beta}_{n}^0g_n(\lambda_{e}r)\right]S_{n}^0(\theta,\phi)e^{\lambda_e^2 t},
%\end{tabular*}
%\end{split}
\label{eq:23}
\end{eqnarray}
\begin{eqnarray}
\eta^e_0=\sum\limits_{n=1}^\infty \left[\gamma_{n}^0f_n(\lambda_{e}r)+\hat{\gamma}_{n}^0g_n(\lambda_{e}r)\right]T_{n}^0(\theta,\phi)e^{\lambda_e^2 t},
\label{eq:24}
\end{eqnarray}
where $\bar{\alpha}_{n}^0$, $\bar{\beta}_{n}^0$, $\bar{\gamma}_{n}^0$, $\hat{\alpha}_{n}^0$,
$\hat{\beta}_{n}^0$, $\hat{\gamma}_{n}^0$  are the unknown coefficients which
are to be determined subject to the boundary conditions
(\ref{eq:14}) to (\ref{eq:18}), and $\lambda_i$, $\lambda_e$ are
the amplification factors corresponding to the flow inside and
outside of the drop which can be found if the initial conditions are provided (Ref. \cite{venkatalaxmi2004general}). Moreover, the far field condition turns out to be $\chi^e_0 \rightarrow \chi_0^\infty$ and
$\eta^e_0 \rightarrow \eta_0^\infty$ as $r\rightarrow\infty$. The unknown
coefficients can be expressed in terms of the known ambient flow
variables using the boundary conditions. We present these details
in Appendix A.\\ \\
The zeroth order drag force experienced by a spherical drop can be computed using the formula
\begin{equation}
\vec{D}=\int_{\theta=0}^{\pi} \int_{\phi=0}^{2\pi} \bar{\bar{\tau}}.\hat{n}\, dS,
\end{equation}
where $dS$ represents the surface element, $\hat{n}$ is the unit
normal to the boundary of the drop, $\textbf{r}$ is the position
vector and $\bar{\bar{\tau}}$ is the stress tensor. We have
computed zeroth order thermocapillary drift in case of transient
Stokes flow past a viscous drop, and expressed in terms of Faxen's laws, given by
\begin{eqnarray}
\vec{D}_0=4\pi \lambda_{e}^{2}\hat{\alpha}_{1}^0\left(A_{11}^0\hat{i}+B_{11}^0\hat{j}+A_{10}^0\hat{k}\right)e^{\lambda_e^2 t}.
\label{eq:27}
\end{eqnarray}
Note that the above structure in terms of the known vector
$(A_{11}^0,\,B_{11}^0,\,A_{10}^0) $ is
due to the spherical harmonics $S_n^0(\theta,\phi)$ given in (\ref{eq:h1}). Corresponding to a given ambient flow, one can determine the coefficient $\hat{\alpha}_{1}^0$. For
example, in case of uniform ambient flow, we have $n=1$ and the corresponding expression for $\hat{\alpha}_{1}^0$ can be obtained using $\hat{\alpha}_{n}^0$ given in Appendix A. Consequently from Eq. (\ref{eq:27}), we have the following
expression for the drag force
\begin{eqnarray}
%\begin{split}
\vec{D}_0=2\pi\left[\frac{Y+\mu X+\alpha P}{W+\mu Z+\alpha G}[\vec{{u}}_{0\infty}]_0+\frac{V+\mu U+\alpha H}{W+\mu Z+\alpha G}[\vec{\nabla}^2 \vec{{u}}_{0\infty}]_0\right].
 %\end{split}
 \label{eq:29}
\end{eqnarray}
The above quantity depends on
$\mu=\frac{\mu^i}{\mu^e}$, the ratio of the viscosities, and $\alpha$ the dimensionless slip coefficient. Since $\Gamma_0=1$, $\vec{\nabla}_s \Gamma_0$ vanishes and the tangential stress becomes continuous. Hence at leading order, we do not observe any influence of the surfactant. The
expanded form of the quantities $X,Y,P,G,Z,W,U,V,H$ etc., are given in
Appendix B. It may
be noted that the above compact form is due to the following
relations
\begin{eqnarray}
[\vec{{u}}_{0\infty}]_0=(2\alpha_{1}^0+\frac{2}{3}\lambda_e
\beta_{1}^0)(A_{11}^0\hat{i}+B_{11}^0\hat{j}+A_{10}^0\hat{k})e^{\lambda_e^2 t},\nonumber
\end{eqnarray}
\begin{eqnarray}
[\vec{\nabla}^2\vec{{u}}_{0\infty}]_0=\frac{2}{3}\lambda_e^3
\beta_{1}^0(A_{11}^0\hat{i}+B_{11}^0\hat{j}+A_{10}^0\hat{k})e^{\lambda_e^2 t},\nonumber
\end{eqnarray}
\begin{eqnarray}
[\vec{\nabla}\times \vec{{u}}_{0\infty}]_0=\frac{2\lambda_e}{3}\gamma_{1}^0(C_{11}^0\hat{i}+D_{11}^0\hat{j}+C_{10}^0\hat{k})e^{\lambda_e^2 t}.\nonumber
\end{eqnarray}
One may observe that, when the slip coefficient in the zeroth order drag force is equal to zero (i.e., $\alpha=0$), then the drag force reduces to
\begin{eqnarray}
\vec{D}_0=2\pi\left[\frac{Y+\mu X}{W+\mu Z}[\vec{{u}}_{0\infty}]_0+\frac{V+\mu U}{W+\mu Z}[\vec{\nabla}^2 \vec{{u}}_{0\infty}]_0\right].
 %\end{split}
 \label{eq:29com1}
\end{eqnarray}
In the context of thermocapillary migration of a spherical drop, Sharanya and Raja Sekhar \cite{sharanya2015thermocapillary} obtained an expression for the drag force exerted on the spherical drop. The above expression (\ref{eq:29com1}) agrees with their results when the thermocapillary effects are neglected. Table (\ref{table1}) gives some additional understanding in this regard.
Note that the zeroth order drag force given in (\ref{eq:29}) is with respect to a reference frame which is
moving with a velocity $\textsl{\textbf{U}}_0$. Therefore the drag force
in the laboratory reference frame in terms of a given ambient
hydrodynamic field is given by
\begin{eqnarray}
\vec{D}_0=2\pi\left[\frac{Y+\mu X+\alpha P}{W+\mu Z+\alpha G}\left([\vec{{v}}_{0\infty}]_0-\textbf{U}_0\right)+\frac{V+\mu U+\alpha H}{W+\mu Z+\alpha G}[\vec{\nabla}^2 \vec{{v}}_{0\infty}]_0\right],
 \label{eq:r29}
\end{eqnarray}
where $\textbf{U}_0$ is the zeroth order migration velocity which is yet to be determined.\\ \\
The force balance in the
absence of gravity when the flow is transient is given by
(Refs.~\cite{subramanian2001motion,chisnell1987unsteady}),
\begin{equation}
M\frac{d \textbf{U}}{dt}=\vec{D},
\label{eq:force}
\end{equation}
where $M=\frac{4}{3}\pi \rho_i$ is the mass of the drop with unit radius. Here, $\rho_i$ is the density of the drop.
From the above equation (\ref{eq:force}), we have the leading order force balance as follows
\begin{equation}
M\frac{d \textbf{U}_0 }{dt}=\vec{D_0}\,.
\end{equation}
On using the expression for the drag given in (\ref{eq:r29}) (general case), this
would enable us to obtain the following expression for the
migration velocity of the drop
\begin{eqnarray}
\textbf{U}_0 & = & \frac{3}{2\rho_i+\rho_e}\left[\frac{Y+\mu X+\alpha P}{W+\mu
Z+\alpha G}[\vec{{v}}_{0\infty}]_0+\frac{V+\mu U+\alpha H}{W+\mu Z+\alpha G}[\nabla^2
\vec{{v}}_{0\infty}]_0 \right] \nonumber\\
&& \left(\frac{3}{2\rho_i+\rho_e}\frac{Y+\mu
X+\alpha P}{W+\mu Z+\alpha G}+\lambda_e^2\right)^{-1}.
 \label{eq:m1}
\end{eqnarray}
%
%\begin{dmath}
%\textbf{U}_0=\frac{3}{2\rho_i+\rho_e}\left[\frac{Y+\mu X+\alpha P}{W+\mu
%Z+\alpha G}[\vec{{v}}_{0\infty}]_0+\frac{V+\mu U+\alpha H}{W+\mu Z+\alpha G}[\nabla^2
%\vec{{v}}_{0\infty}]_0 \right]\left(\frac{3}{2\rho_i+\rho_e}\frac{Y+\mu
%X+\alpha P}{W+\mu Z+\alpha G}+\lambda_e^2\right)^{-1}.
% \label{eq:m1}
%\end{dmath}
We may observe that, when the slip coefficient is zero, the above zeroth order migration velocity reduces to the one that is obtained by Sharanya and Raja Sekhar \cite{sharanya2015thermocapillary} provided thermal effects are neglected. In this case, we have
\begin{eqnarray}
\textbf{U}_0 & = & \frac{3}{2\rho_i+\rho_e}\left[\frac{Y+\mu X}{W+\mu
Z}[\vec{{v}}_{0\infty}]_0+\frac{V+\mu U}{W+\mu Z}[\nabla^2
\vec{{v}}_{0\infty}]_0 \right] \nonumber\\
&& \left(\frac{3}{2\rho_i+\rho_e}\frac{Y+\mu
X}{W+\mu Z}+\lambda_e^2\right)^{-1}.
 \label{eq:m1comp1}
\end{eqnarray}
If we consider the limiting case of no oscillations in the hydrodynamic flow field, i.e., $\lambda_i=\lambda_e=0$, and zero slip coefficient, i.e., $\alpha=0$, then the zeroth order terminal velocity reduces to
\begin{equation}
\textbf{U}_0=[\vec{{v}}_\infty]_0+\frac{\mu}{4+6\mu}[\vec{\nabla}^2
\vec{{v}}_\infty]_0,
\label{eq:m2comp1}
\end{equation}
which is exactly matching with the one that is obtained by Pak, Feng and Stone \cite{pak2014viscous}.
\subsubsection{Stationary drop}
If we assume that the drop is stationary, then we have $\vec{{v}}_\infty=\vec{{u}}_\infty$. In this case, the zeroth order drag force is given by
\begin{eqnarray}
\vec{D}_0=2\pi\left[\frac{Y+\mu X+\alpha P}{W+\mu Z+\alpha G}[\vec{{v}}_{0\infty}]_0+\frac{V+\mu U+\alpha H}{W+\mu Z+\alpha G}[\vec{\nabla}^2 \vec{{v}}_{0\infty}]_0\right],
 \label{eq:r29ch}
\end{eqnarray}
which agrees with the corresponding result that is obtained by Choudhuri and Padmavati \cite{choudhuri2014study} when the slip coefficient is zero (Ref. Table (\ref{table1})).
\subsection{\label{firstorder}First-order correction}
The first order surfactant transport equation due to the expansion (\ref{eq:expansion}) and Eq. (\ref{eq:diffusiona}) is given by
\begin{eqnarray}
Pr_s\frac{\partial \Gamma_1}{\partial t}+\vec{\nabla}_s.\vec{u}_{0s}=\vec{\nabla}_s^2 \Gamma_1,
\label{eq:diffusionc}
\end{eqnarray}
where $\vec{u}_{0s}$ is the zeroth order tangential velocity vector on the drop surface. Assuming that the surfactant concentration is oscillatory, i.e., $\Gamma_1(\theta,\phi,t)=\Gamma_1(\theta,\phi)e^{-i\omega t}=\Gamma_1(\theta,\phi)e^{-l^2 t/Pr_s}$, Eq. (\ref{eq:diffusionc}) reduces to
\begin{eqnarray}
(\vec{\nabla}_s^2+l^2)\Gamma_1= \vec{\nabla}_s.\vec{u}_{0s}.
\label{eq:diffusiond}
\end{eqnarray}
In order to obtain the first order surfactant concentration $\Gamma_1$, we express $\Gamma_1$ in terms of spherical harmonics, i.e.,
\begin{eqnarray}
\Gamma_1=\sum\limits_{n=1}^\infty R_{n}^1(\theta,\phi)e^{-l^2 t/Pr_s},
\label{eq:diffusione}
\end{eqnarray}
where
\begin{eqnarray}
R_{n}^1(\theta,\phi)=\sum\limits_{m=0}^n
\left(E_{nm}^1\cos\,m\phi+F_{nm}^1\sin\,m\phi\right)P_n^m(\cos\,\theta),
\end{eqnarray}
are the spherical harmonics, and $E_{nm}^1$, $F_{nm}^1$ have to be determined such that $\Gamma_1$ satisfies the Eq.(\ref{eq:diffusione}). Since $\vec{\nabla}_s^2 R_{n}^1(\theta,\phi)=-n(n+1)R_{n}^1(\theta,\phi)$, we observe that $\vec{\nabla}_s^2 \Gamma_1 =-\sum\limits_{n=1}^\infty n(n+1)R_{n}^1(\theta,\phi)e^{l^2 t/Pr_s}$. The coefficients in $R_{n}^1(\theta,\phi)$ can be determined as follows:
\begin{eqnarray}
 \sum\limits_{n=0}^\infty \sum\limits_{m=0}^n (-n(n+1)+l^2)\left[E_{nm}^1 \cos\,m \phi+F_{nm}^1 \sin\,m \phi\right]P_n^m(\cos\,\theta)e^{-l^2t/Pr_s} =
 \vec{\nabla}_s.\vec{u}_{0s}. \nonumber \\
 \end{eqnarray}
%
%\begin{dmath}
% \sum\limits_{n=1}^\infty \sum\limits_{m=0}^n (-n(n+1)+l^2)\left[E_{nm}(\psi) \cos\,m \phi+F_{nm}(\psi) \sin\,m \phi\right]P_n^m(\cos\,\theta)e^{l^2t}=\begin{cases}
% \vec{\nabla}_s.\vec{u}_{0s} & \mbox{ for $0\leq\theta\leq\psi$}, \\ 0 & \mbox{ for $\psi<\theta\leq\pi$}.
% \end{cases}
%\end{dmath}
 This enables us to write the following relations
\begin{eqnarray}
 E_{kj}^1\pi\frac{2(k+j)!}{(2k+1)(k-j)!}e^{-l^2t/Pr_s}&=&
 \frac{-1}{k(k+1)-l^2}\int_{\phi=0}^{2\pi}\int_{\theta=0}^{\pi}
 (\vec{\nabla}_s.\vec{u}_{0s})P_k^j(\cos\,\theta)\nonumber\\
 && \cos\,j\phi\sin\,\theta\,d\theta\,d\phi,
\end{eqnarray}
\begin{eqnarray}
 F_{kj}^1\pi\frac{2(k+j)!}{(2k+1)(k-j)!}e^{-l^2t/Pr_s}&=&
 \frac{-1}{k(k+1)-l^2}\int_{\phi=0}^{2\pi}\int_{\theta=0}^{\pi}
 (\vec{\nabla}_s.\vec{u}_{0s})P_k^j(\cos\,\theta)\nonumber\\
 && \sin\,j\phi\sin\,\theta\,d\theta\,d\phi,
\end{eqnarray}
which implies $-l^2/Pr_s=\lambda_e^2(<0)$ and
\begin{eqnarray}
 E_{nm}^1& = &\left[(n+1)\alpha_{n}^0+\beta_{n}^0
 \left(\lambda_ef_{n+1}(\lambda_e)+(n+1)f_n(\lambda_e)\right)\right. \nonumber\\
  && \left. \mbox{} -n\hat{\alpha}_{n}^0+\hat{\beta}_{n}^0
 \left((n+1)g_n(\lambda_e)-\lambda_eg_{n+1}(\lambda_e)\right)\right]
 \times\left[A_{nm}^0\frac{n(n+1)}{n(n+1)+\lambda_e^2Pr_s}\right], \nonumber\\
 \label{enm}
\end{eqnarray}
\begin{eqnarray}
 F_{nm}^1& = &\left[(n+1)\alpha_{n}^0+\beta_{n}^0
 \left(\lambda_ef_{n+1}(\lambda_e)+(n+1)f_n(\lambda_e)\right)\right. \nonumber\\
  && \left. \mbox{} -n\hat{\alpha}_{n}^0+\hat{\beta}_{n}^0
 \left((n+1)g_n(\lambda_e)-\lambda_eg_{n+1}(\lambda_e)\right)\right]
 \times\left[B_{nm}^0\frac{n(n+1)}{n(n+1)+\lambda_e^2Pr_s}\right]. \nonumber\\
 \label{fnm}
\end{eqnarray}

The first-order pressure and velocity fields satisfy the unsteady Stokes and continuity equations. Correspondingly, we express  $\chi^{i}_1$, $\eta^{i}_1$, $\chi^{e}_1$ and $\eta^e_1$ as follows
\begin{eqnarray}
\chi^{i}_1=\sum\limits_{n=1}^\infty \left[\bar{\alpha}_{n}^1r^{n}+\bar{\beta}_{n}^1f_n(\lambda_{i}r)\right]S_{n}^1(\theta,\phi)e^{\lambda_i^2 t},
\label{eq:21f}
\end{eqnarray}
\begin{eqnarray}
\eta^{i}_1=\sum\limits_{n=1}^\infty \left[\bar{\gamma}_{n}^1f_n(\lambda_{i}r)\right]T_{n}^1(\theta,\phi)e^{\lambda_i^2 t},
\label{eq:22f}
\end{eqnarray}
\begin{eqnarray}
%\begin{split}
\chi^{e}_1=\sum\limits_{n=1}^\infty \left[\alpha_{n}^1r^{n}+\frac{\hat{\alpha}_{n}^1}{r^{n+1}}+\beta_{n}^1f_n(\lambda_{e}r)+\hat{\beta}_{n}^1g_n(\lambda_{e}r)\right]S_{n}^1(\theta,\phi)e^{\lambda_e^2 t},
%\end{tabular*}
%\end{split}
\label{eq:23f}
\end{eqnarray}
\begin{eqnarray}
\eta^{e}_1=\sum\limits_{n=1}^\infty \left[\gamma_{n}^1f_n(\lambda_{e}r)+\hat{\gamma}_{n}^1g_n(\lambda_{e}r)\right]T_{n}^1(\theta,\phi)e^{\lambda_e^2 t},
\label{eq:24f}
\end{eqnarray}
where $S_{n}^1(\theta,\phi)$ and $T_{n}^1(\theta,\phi)$ are spherical harmonics of order $n$. The interfacial surfactant that is coupled via the boundary conditions (\ref{eq:16a}) and (\ref{eq:16b}) together with the form of $\Gamma_1$ given in (\ref{eq:diffusione}) enforces $S_{n}^1(\theta,\phi)=R_{n}^1(\theta,\phi)$. However, we have
\begin{eqnarray}
T_{n}^1(\theta,\phi)=\sum\limits_{m=0}^n
\left(E_{nm}'\cos\,m\phi+F_{nm}'\sin\,m\phi\right)P_n^m(\cos\,\theta).
\end{eqnarray}
We have given the expressions for the unknown coefficients, $\alpha_{n}^1$, $\hat{\alpha}_{n}^1$, $\beta_{n}^1$, $\hat{\beta}_{n}^1$, $\gamma_{n}^1$, $\hat{\gamma}_{n}^1$, $\bar{\alpha}_{n}^1$, $\bar{\beta}_{n}^1$ and $\bar{\gamma}_{n}^1$, in Appendix C.
Following a similar approach that is used to solve the leading order problem, we compute the first order drag given by
\begin{eqnarray}
\vec{D}_1 & = & 2\pi\Big[-\frac{Y+\mu X+\alpha P}{W+\mu Z+\alpha G}{\textbf{U}}_1+\frac{2 Ma \lambda_e^2f_2(\lambda_i)g_1(\lambda_e)}{(W+\mu Z+\alpha G)}  \nonumber\\
&&  \mbox{}\times (E_{11}^1\hat{i}+F_{11}^1\hat{j}+E_{10}^1\hat{k})e^{\lambda_e^2 t}\Big].
 \label{eq:r29f}
\end{eqnarray}
The force balance $M\frac{d {\textbf{U}}_1}{dt}=\vec{D}_1$ together with the expression for $\vec{D}_1$ given in Eq. (\ref{eq:r29f}) leads to the first order
migration velocity of the drop
\begin{eqnarray}
{\textbf{U}}_1 & = & \frac{3}{2\rho_i+\rho_e}\left[\frac{2 Ma \lambda_e^2f_2(\lambda_i)g_1(\lambda_e)}{(W+\mu Z+\alpha G)}(E_{11}^1\hat{i}+F_{11}^1\hat{j}+E_{10}^1\hat{k})e^{\lambda_e^2 t}\right]\nonumber\\
&& \mbox{}\times \left(\frac{3}{2\rho_i+\rho_e}\frac{Y+\mu
X+\alpha P}{W+\mu Z+\alpha G}+\lambda_e^2\right)^{-1},
\label{eq:m1f}
\end{eqnarray}
where
\begin{eqnarray}
E_{11}^1=\frac{2 A_{11}^0}{(2+\lambda_e^2Pr_s)}\left[2\alpha_{n}^0+\beta_{n}^0
 \left(\lambda_ef_{2}(\lambda_e)+2f_1(\lambda_e)\right)-\hat{\alpha}_{n}^0+\hat{\beta}_{n}^0
 \left(2g_1(\lambda_e)-\lambda_eg_{2}(\lambda_e)\right)\right], \nonumber\\
 \label{eq:ab2}
\end{eqnarray}
\begin{eqnarray}
F_{11}^1=\frac{2 B_{10}^0}{(2+\lambda_e^2Pr_s)}\left[2\alpha_{n}^0+\beta_{n}^0
 \left(\lambda_ef_{2}(\lambda_e)+2f_1(\lambda_e)\right)-\hat{\alpha}_{n}^0+\hat{\beta}_{n}^0
 \left(2g_1(\lambda_e)-\lambda_eg_{2}(\lambda_e)\right)\right], \nonumber\\
 \label{eq:ab3}
\end{eqnarray}
and
\begin{eqnarray}
E_{10}^1=\frac{2 A_{10}^0}{(2+\lambda_e^2Pr_s)}\left[2\alpha_{n}^0+\beta_{n}^0
 \left(\lambda_ef_{2}(\lambda_e)+2f_1(\lambda_e)\right)-\hat{\alpha}_{n}^0+\hat{\beta}_{n}^0
 \left(2g_1(\lambda_e)-\lambda_eg_{2}(\lambda_e)\right)\right], \nonumber\\
 \label{eq:ab1}
\end{eqnarray}
Here we observe that, only three modes of concentration $E_{11}^1$, $ F_{11}^1$ and $ E_{10}^1$ are contributing to the drag and migration velocity.
If we consider the special case of steady flow past a droplet, i.e., $\lambda_e=\lambda_i=0$, the first order migration velocity reduces to
\begin{eqnarray}
\textbf{U}_1=\frac{2 Ma}{6+9 \mu+18 \alpha \mu}(e_{11}^1\hat{i}+f_{11}^1\hat{j}+e_{10}^1\hat{k}),
\label{eq:m1fcompsi}
\end{eqnarray}
where
\begin{eqnarray}
e_{kj}^1\pi\frac{2(k+j)!}{(2k+1)(k-j)!}=\frac{-1}{k(k+1)}\int_{\phi=0}^{2\pi}\int_{\theta=0}^{\pi}
 (\vec{\nabla}_s.\vec{u}_{0s})P_k^j(\cos\,\theta)\cos\,j\phi\sin\,\theta\,d\theta\,d\phi,\nonumber\\
 \end{eqnarray}
\begin{eqnarray}
 f_{kj}^1\pi\frac{2(k+j)!}{(2k+1)(k-j)!}=\frac{-1}{k(k+1)}\int_{\phi=0}^{2\pi}\int_{\theta=0}^{\pi}
 (\vec{\nabla}_s.\vec{u}_{0s})P_k^j(\cos\,\theta)\sin\,j\phi\sin\,\theta\,d\theta\,d\phi.\nonumber\\
\end{eqnarray}
In particular,
\begin{eqnarray}
e_{11}^1={A_{11}^0}\left[\frac{\alpha_{1}^0 \left(1+3 \alpha  \mu\right)}{1+\mu +3 \alpha  \mu}\right],
 \label{eq:ab2compsi}
\end{eqnarray}
\begin{eqnarray}
f_{11}^1={B_{10}^0}\left[\frac{\alpha_{1}^0 \left(1+3 \alpha  \mu\right)}{1+\mu +3 \alpha  \mu}\right],
 \label{eq:ab3compsi}
\end{eqnarray}
and
\begin{eqnarray}
e_{10}^1={A_{10}^0}\left[\frac{\alpha_{1}^0 \left(1+3 \alpha  \mu\right)}{1+\mu +3 \alpha  \mu}\right].
 \label{eq:ab1compsi}
\end{eqnarray}
If the slip coefficient $\alpha=0$, this result is matching with the one obtained by Pak, Feng, Stone \cite{pak2014viscous}.
\subsubsection{Stationary drop}
If we assume that the drop is stationary, the first order drag force is given by
\begin{eqnarray}
\vec{D}_1=2\pi\left[\frac{2 Ma \lambda_e^2f_2(\lambda_i)g_1(\lambda_e)}{(W+\mu Z+\alpha G)}(E_{11}^1\hat{i}+F_{11}^1\hat{j}+E_{10}^1\hat{k})e^{\lambda_e^2 t}\right].
\end{eqnarray}

\subsection{\label{secondorder}Second-order correction}
The second order surfactant transport equation is given by
\begin{eqnarray}
Pr_s\frac{\partial \Gamma_2}{\partial t}+\vec{\nabla}_s.(\Gamma_0\vec{u}_{1s}+\Gamma_1\vec{u}_{0s})=\vec{\nabla}_s^2 \Gamma_1,
\label{eq:diffuc}
\end{eqnarray}
where $\vec{u}_{0s}$, $\vec{u}_{1s}$ are the zeroth order and first order tangential velocity components on the drop surface respectively. Assuming that the surfactant concentration is oscillatory, i.e., $\Gamma_2(\theta,\phi,t)=\Gamma_2(\theta,\phi)e^{-i\omega_2 t}=\Gamma_2(\theta,\phi)e^{-l_2^2 t/Pr_s}$, Eq. (\ref{eq:diffuc}) reduces to
\begin{eqnarray}
(\vec{\nabla}_s^2+l_2^2)\Gamma_2=  \vec{\nabla}_s.(\Gamma_0\vec{u}_{1s}+\Gamma_1\vec{u}_{0s}).
\label{eq:diffud}
\end{eqnarray}
In order to obtain the second order surfactant concentration,  $\Gamma_2$, we adopt a similar procedure that is used in Section. (\ref{firstorder}).  We express $\Gamma_2$ in terms of spherical harmonics, i.e.,
\begin{eqnarray}
\Gamma_2=\sum\limits_{n=1}^\infty R_{n}^2(\theta,\phi)e^{-l_2^2 t/Pr_s},
\label{eq:diffue}
\end{eqnarray}
where
\begin{eqnarray}
R_{n}^2(\theta,\phi)=\sum\limits_{m=0}^n
\left(E_{nm}^2\cos\,m\phi+F_{nm}^2\sin\,m\phi\right)P_n^m(\cos\,\theta),
\end{eqnarray}
and $E_{nm}^2$, $F_{nm}^2$ have to be determined such that $\Gamma_2$ satisfies the Eq.(\ref{eq:diffud}). Correspondingly, the coefficients $R_{n}^2(\theta,\phi)$ can be determined as follows:
\begin{eqnarray}
 E_{kj}^2\pi\frac{2(k+j)!}{(2k+1)(k-j)!}e^{-l_2^2t/Pr_s}=
 \frac{-1}{k(k+1)-l_2^2}\int_{\phi=0}^{2\pi}\int_{\theta=0}^{\pi}
 (\vec{\nabla}_s.(\Gamma_0\vec{u}_{1s}+\Gamma_1\vec{u}_{0s}))
 \nonumber\\ P_k^j(\cos\,\theta)\cos\,j\phi\sin\,\theta\,d\theta\,d\phi, \nonumber \\
\end{eqnarray}
\begin{eqnarray}
 F_{kj}^2\pi\frac{2(k+j)!}{(2k+1)(k-j)!}e^{-l_2^2t/Pr_s}=
 \frac{-1}{k(k+1)-l_2^2}\int_{\phi=0}^{2\pi}\int_{\theta=0}^{\pi}
 (\vec{\nabla}_s.(\Gamma_0\vec{u}_{1s}+\Gamma_1\vec{u}_{0s}))
 \nonumber\\  P_k^j(\cos\,\theta)\sin\,j\phi\sin\,\theta\,d\theta\,d\phi, \nonumber \\
\end{eqnarray}
which implies $-l_2^2/Pr_s=\lambda_e^2$, and
\begin{eqnarray}
 E_{kj}^2& = &\left[-k \hat{\alpha}_{k}^2+\hat{\beta}_{k}^2
 \left((k+1)g_k(\lambda_e)-\lambda_eg_{k+1}(\lambda_e)\right)\right]\left[ E_{kj}^1\frac{k(k+1)}{k(k+1)+\lambda_e^2Pr_s}\right]
 -\frac{(2k+1)(k-j)!}{2\pi(k+j)!} \nonumber\\ && \mbox{} \times\frac{e^{-\lambda_e^2t}}{k(k+1)+\lambda_e^2Pr_s}\int_{\phi=0}^{2\pi}\int_{\theta=0}^{\pi}
 (\vec{\nabla}_s.(\Gamma_1\vec{u}_{0s}))P_k^j(\cos\,\theta)
 \cos\,j\phi\sin\,\theta\,d\theta\,d\phi,
 \label{inm}
\end{eqnarray}
\begin{eqnarray}
 F_{kj}^2& = &\left[-k \hat{\alpha}_{k}^2+\hat{\beta}_{k}^2
 \left((k+1)g_k(\lambda_e)-\lambda_eg_{k+1}(\lambda_e)\right)\right]\left[ F_{kj}^1\frac{k(k+1)}{k(k+1)+\lambda_e^2Pr_s}\right]
 -\frac{(2k+1)(k-j)!}{2\pi(k+j)!} \nonumber\\ && \mbox{} \times\frac{e^{-\lambda_e^2t}}{k(k+1)+\lambda_e^2Pr_s}\int_{\phi=0}^{2\pi}\int_{\theta=0}^{\pi}
 (\vec{\nabla}_s.(\Gamma_1\vec{u}_{0s}))P_k^j(\cos\,\theta)
 \sin\,j\phi\sin\,\theta\,d\theta\,d\phi.
 \label{jnm}
\end{eqnarray}
Evaluating the double integral on the right hand side is difficult for any given arbitrary flow. However, these can be evaluated for a given ambient flow so that we have the second order concentration. Accordingly, we compute these double integrals for specific cases like uniform flow, Couette flow etc.

Once we obtain the second order concentration for a given flow, one can solve the above equations by following similar procedure that is used to solve the zeroth and first order equations. The second order drag is given by
\begin{eqnarray}
\vec{D}_2& = &2\pi\left[\frac{Y+\mu X+\alpha P}{W+\mu Z+\alpha G}\left(-{\textbf{U}}_2\right)+\frac{2 Ma \lambda_e^2f_2(\lambda_i)g_1(\lambda_e)}{(W+\mu Z+\alpha G)}\right. \nonumber\\
  && \left. \mbox{} \times (E_{11}^2\hat{i}+F_{11}^2\hat{j}+E_{10}^2\hat{k})e^{\lambda_e^2 t}\right].
 \label{eq:r29s}
\end{eqnarray}
The force balance $M\frac{d \textbf{U}_2}{dt}=\vec{D_2}$ leads to
%(\ref{eq:force}), for the second order problem reduces to
%\begin{equation}
%M\frac{d \textbf{U}_2}{dt}=\vec{D_2}.
%\end{equation}
%On using the expression for drag given in (\ref{eq:r29s}), we obtain the following expression for the second order
%migration velocity of the drop
\begin{eqnarray}
{\textbf{U}}_2 & = &\frac{3}{2\rho_i+\rho_e}\left[\frac{2 Ma \lambda_e^2f_2(\lambda_i)g_1(\lambda_e)}{(W+\mu Z+\alpha G)}(E_{11}^2\hat{i}+F_{11}^2\hat{j}+E_{10}^2\hat{k})e^{-\lambda_e^2 t}\right]\nonumber\\
  && \mbox{} \left(\frac{3}{2\rho_i+\rho_e}\frac{Y+\mu
X+\alpha P}{W+\mu Z+\alpha G}+\lambda_e^2\right)^{-1},
\label{eq:m1s}
\end{eqnarray}
where $X,Y,P,G,Z,W,U,V,H$ etc., are given in the
Appendix B.
We therefore, conclude that the second order migration velocity and drag depend only on three modes of the concentration namely, $E_{11}^2$,$F_{11}^2$ and $E_{10}^2$.
If we consider the special case of steady flow past a drop, i.e., $\lambda_e=\lambda_i=0$, the second order migration velocity reduces to
\begin{eqnarray}
{\textbf{U}}_2=\frac{2 Ma}{6+9 \mu+18 \alpha \mu_i}(e_{11}^2\hat{i}+f_{11}^2\hat{j}+e_{10}^2\hat{k}),
\label{eq:m1fcomp2}
\end{eqnarray}
where
\begin{eqnarray}
e_{nm}^2=\frac{-1}{n(n+1)}\int_{\phi=0}^{2\pi}\int_{\theta=0}^{\pi}
 (\vec{\nabla}_s.(\Gamma_0\vec{u}_{1s}+\Gamma_1\vec{u}_{0s}))
 P_n^m(\cos\,\theta)\cos\,m\phi\sin\,\theta\,d\theta\,d\phi,\nonumber\\
 \end{eqnarray}
\begin{eqnarray}
 f_{nm}^2=\frac{-1}{n(n+1)}\int_{\phi=0}^{2\pi}\int_{\theta=0}^{\pi}
 (\vec{\nabla}_s.(\Gamma_0\vec{u}_{1s}+\Gamma_1\vec{u}_{0s}))
 P_n^m(\cos\,\theta)\sin\,m\phi\sin\,\theta\,d\theta\,d\phi.\nonumber\\
\end{eqnarray}
If the slip coefficient $\alpha=0$, this result also agrees with the one that is obtained by Pak et al. \cite{pak2014viscous}.
\subsubsection{Stationary drop}
If we assume that the drop is stationary, the second order drag force is given by
\begin{eqnarray}
\vec{D}_2=2\pi\left[\frac{2 Ma \lambda_e^2f_2(\lambda_i)g_1(\lambda_e)}{(W+\mu Z+\alpha G)}(E_{11}^2\hat{i}+F_{11}^2\hat{j}+E_{10}^2\hat{k})e^{\lambda_e^2 t}\right].
\end{eqnarray}

\section{Results and discussion}
Now, we present important observations with reference to some
special cases such as uniform ambient flow, Couette flow, etc.
%\subsection{Uniform flow}
\subsection{Uniform ambient flow}
Consider a uniform flow along the $x-$axis, past a liquid drop of
unit radius whose center is at its origin. In this case,
$\vec{{u}}_{\infty}=\vec{{u}}_{0\infty}=\hat{i}e^{\lambda_e^2
t}$.

\begin{figure}
  \centerline{\includegraphics[width=12cm, height=8cm]{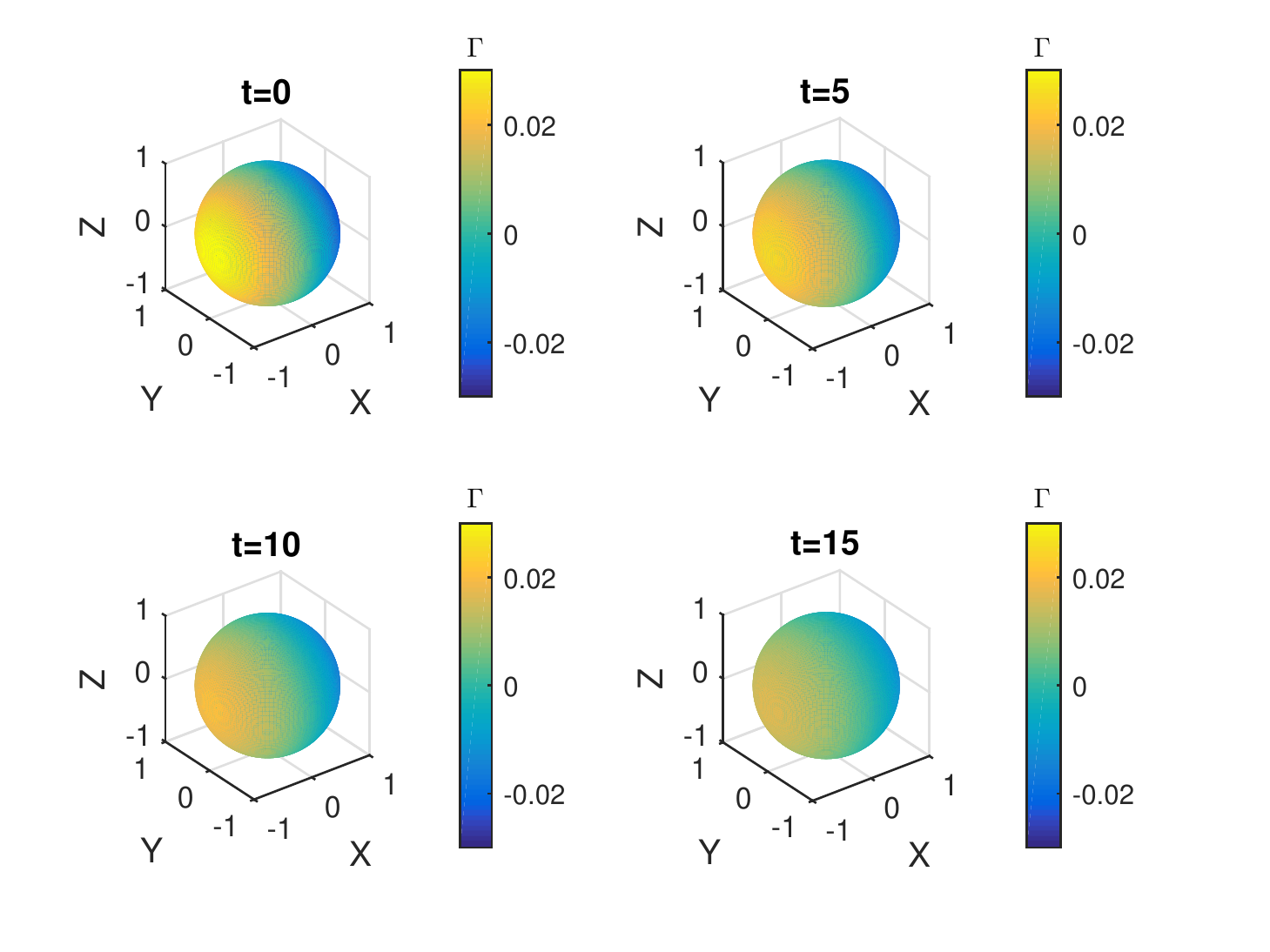}}% Images in 100% size
  \caption{Variation of first order surfactant distribution with the time $t$ corresponding to uniform flow, with $\lambda_e^2=-0.04$, $\lambda_i^2=-0.04$, $\mu=5$ $Ma=400$ and $\alpha=0.1$.}
\label{fig:1e}
\end{figure}
\begin{figure}
  \centerline{\includegraphics[width=12cm, height=8cm]{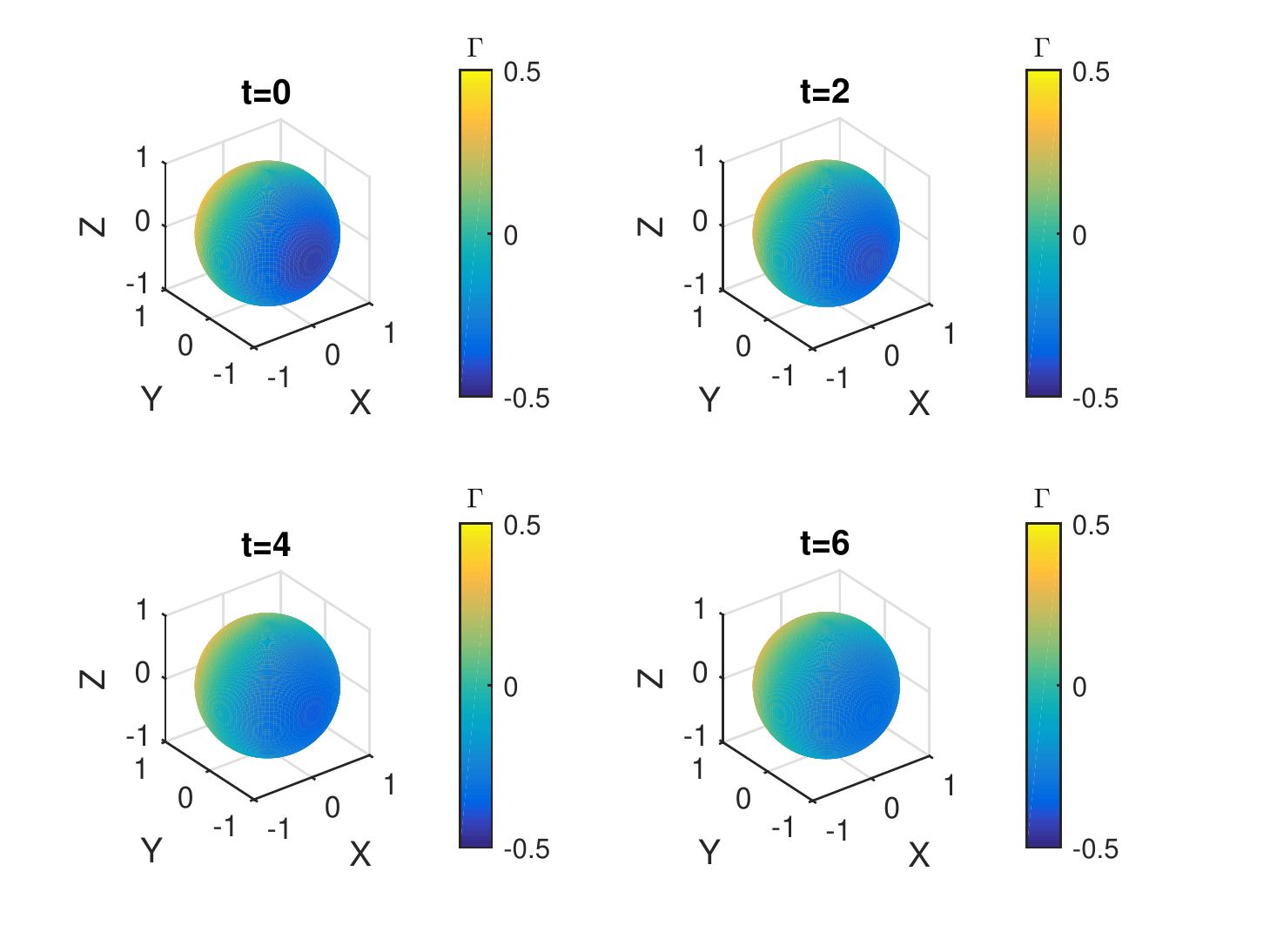}}% Images in 100% size
  \caption{Variation of second order surfactant distribution with the time $t$ corresponding to uniform flow, with $\lambda_e^2=-0.04$, $\lambda_i^2=-0.04$, $\mu=5$ $Ma=400$ and $\alpha=0.1$.}
\label{fig:1f}
\end{figure}
\begin{figure}
  \centerline{\includegraphics[width=7.2cm, height=7.2cm]{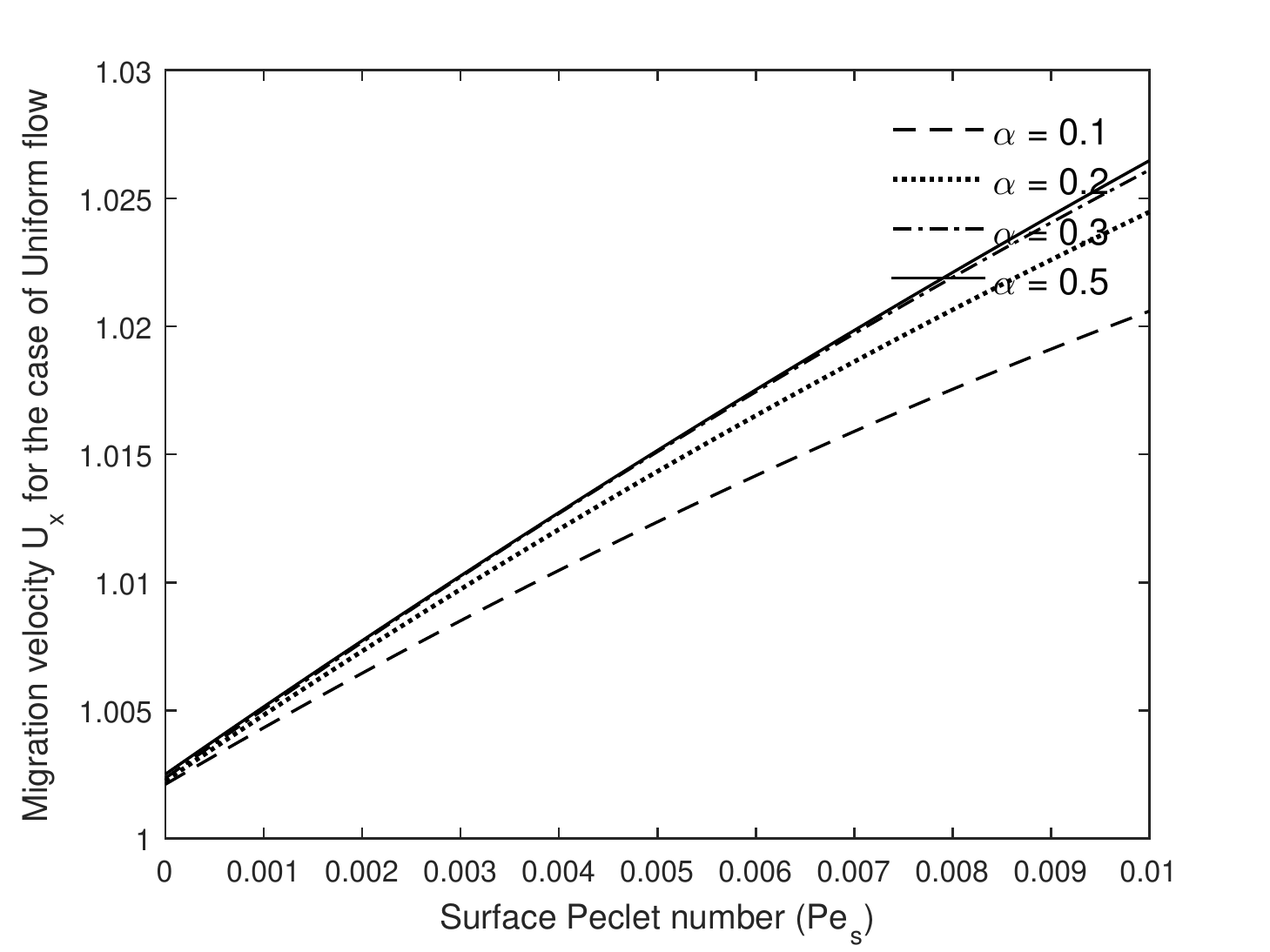}}% Images in 100% size
  \caption{Variation of migration velocity with $Pe_s$ for different slip parameters corresponding to uniform flow, $\alpha$,  $\lambda_e^2=-0.01$, $\lambda_i^2=-0.01$, $Ma=400$ and $\mu=5$.}
\label{fig:8}
\end{figure}

%
%\begin{figure}
%        \centering
%{\includegraphics[width=7.2cm, height=7.2cm]{NUniMuLi}}
%         \caption{Variation of migration velocity with $\lambda_i$, for different viscosity ratios corresponding to uniform flow, $\mu$,  $\lambda_e^2=-0.04$,  $Ma=400$, $\alpha=0.2$ and $Pe_s=0.01$.}
%         \label{fig:10}
%\end{figure}

\begin{figure}
       \begin{subfigure}[b]{0.6\textwidth}
        \centering
{\includegraphics[width=7.2cm, height=7.2cm]{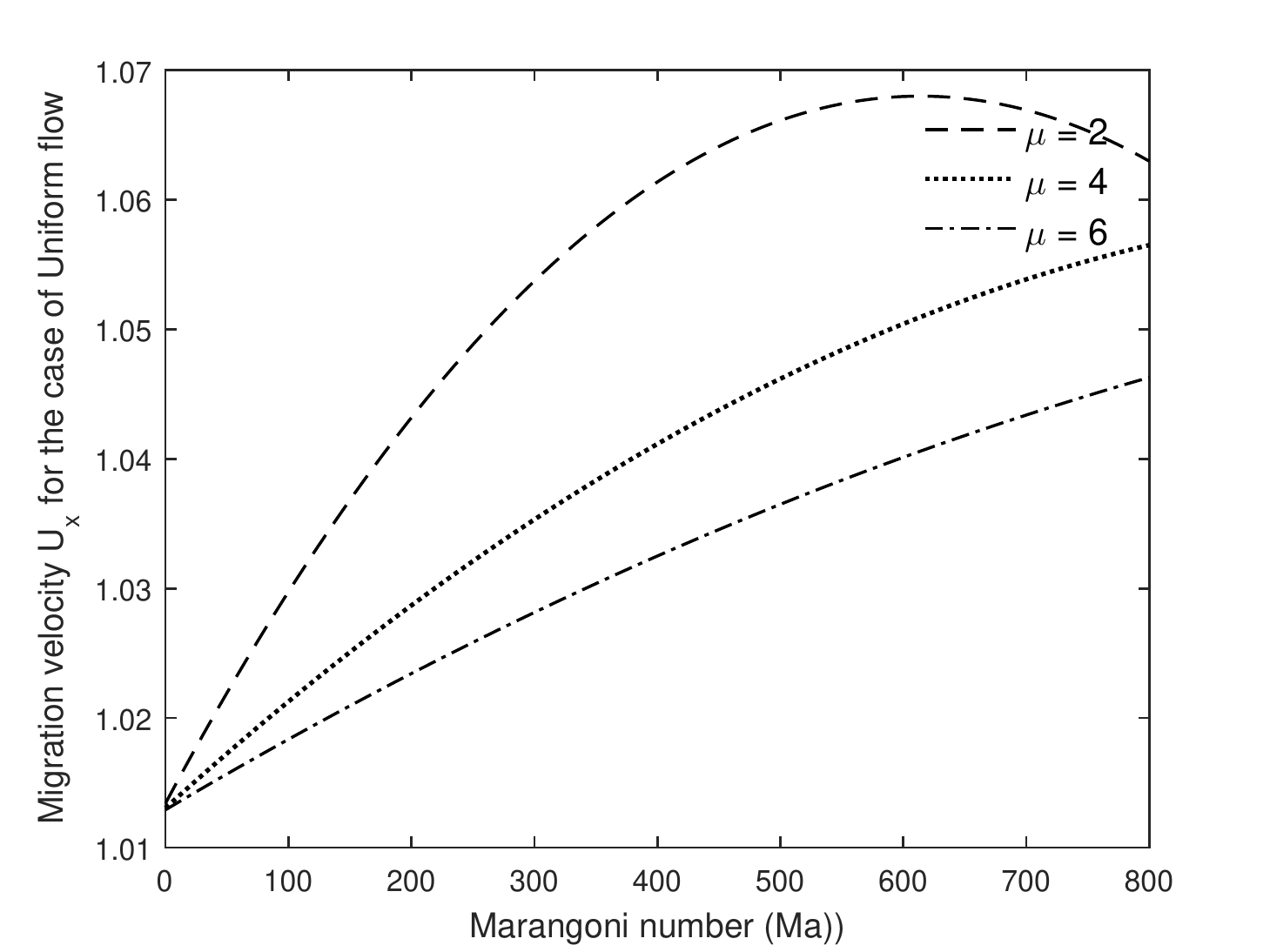}}
%                \caption{A gull}
                %\label{fig:gull}
        \end{subfigure}%
\begin{subfigure}[b]{0.6\textwidth}
               \centering
{\includegraphics[width=7.2cm, height=7.2cm]{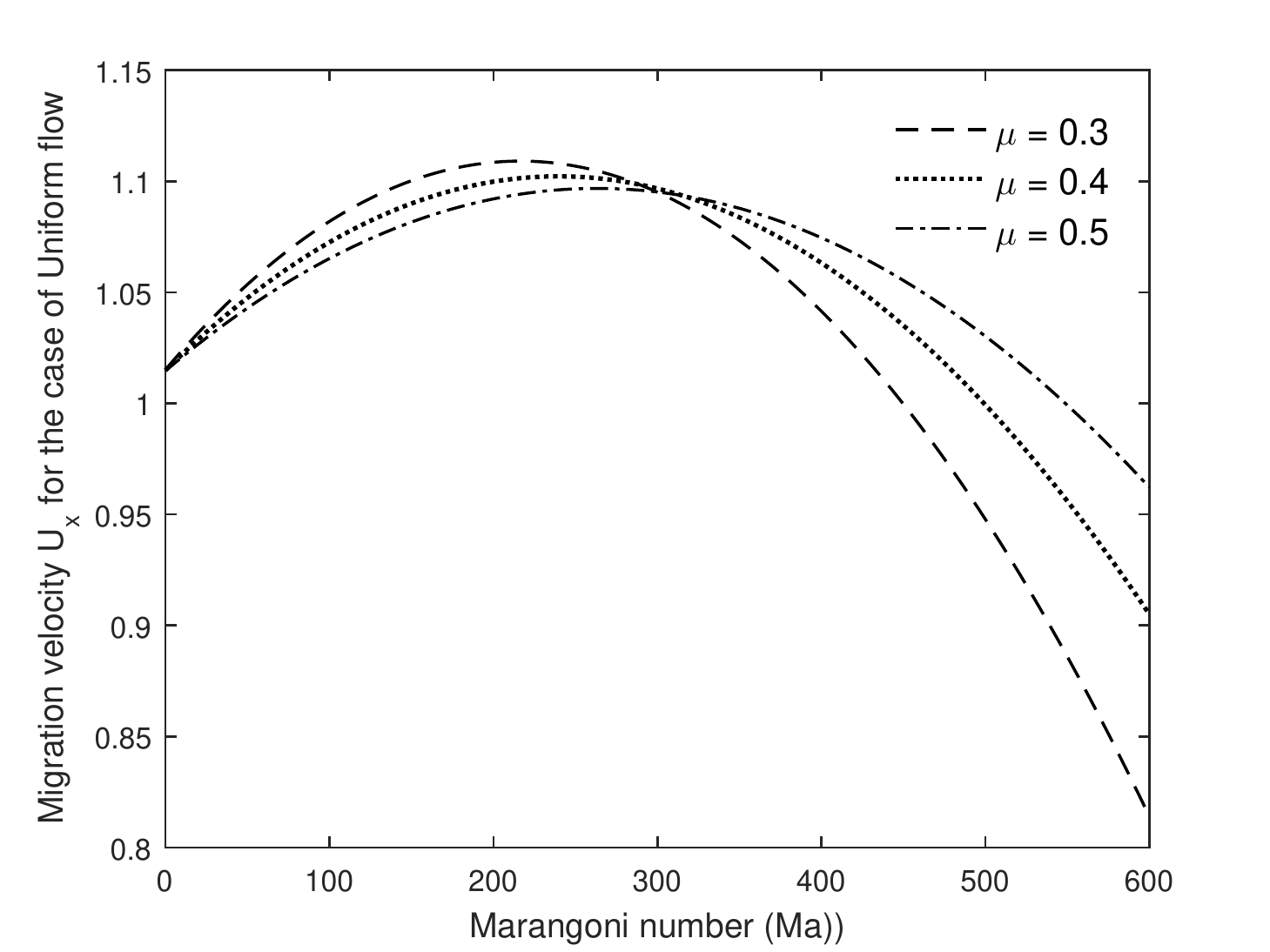}}
%                \caption{A gull2}
               %\label{fig:gull2}
        \end{subfigure}%
         \caption{Variation of migration velocity with Marangoni number ($Ma$) for different viscosity ratios corresponding to uniform flow, $\mu$,  $\lambda_e^2=-0.04$, $\lambda_i^2=-0.04$,  $\alpha=0.2$ and $Pe_s=0.01$.}
         \label{fig:11}
\end{figure}
 Therefore, the corresponding scalar functions $\chi_0^\infty$ and
$\eta_0^\infty$ are given by
\begin{eqnarray}
\chi_0^\infty=\frac{1}{2}r \sin\,\theta \cos\,\phi e^{\lambda_e^2 t},\,\,\,\,\,\eta_0=0.\nonumber
\end{eqnarray}
The above choice indicates that $\alpha_{1}^0=\frac{1}{2}$,
$\beta_{1}^0=0$, $\gamma_{1}^0=0$  in  Eqs.~(\ref{eq:19}) and
(\ref{eq:20}). Therefore
the corresponding drag on the
spherical drop is given by
\begin{eqnarray}
\vec{D}=\vec{D_0}+Pe_s \vec{D_1}+Pe_s^2 \vec{D_2}+O(Pe_s^3),
\label{eq:uni}
\end{eqnarray}
where
\begin{eqnarray}
\vec{D}_0=2\pi\left[\frac{Y+\mu X+\alpha P}{W+\mu Z+\alpha G}\right]e^{\lambda_e^2 t}\hat{i},
\label{eq:uni1}
\end{eqnarray}

\begin{eqnarray}
\vec{D}_1=2\pi\left[\frac{Y+\mu X+\alpha P}{W+\mu Z+\alpha G}\left(-{\textbf{U}}_1\right)+\frac{2 Ma \lambda_e^2f_2(\lambda_i)g_1(\lambda_e)}{(W+\mu Z+\alpha G)}E_{11}^1\hat{i}e^{\lambda_e^2 t}\right],
\label{eq:uni2}
\end{eqnarray}
 \begin{eqnarray}
\vec{D}_2=2\pi\left[\frac{Y+\mu X+\alpha P}{W+\mu Z+\alpha G}\left(-{\textbf{U}}_2\right)+\frac{2 Ma \lambda_e^2f_2(\lambda_i)g_1(\lambda_e)}{(W+\mu Z+\alpha G)}E_{11}^2\hat{i}e^{\lambda_e^2 t}\right].
 \label{eq:uni2g}
\end{eqnarray}
Here
\begin{eqnarray}
E_{11}^1=\frac{2}{(2+\lambda_e^2Pr_s)}\frac{\left(3 g_1(\lambda_e) \lambda _e^2 \left(f_2(\lambda_i)+\alpha  \mu  \left(-2 f_2(\lambda_i)+f_1(\lambda_i) \lambda _i\right) \right)\right)}{\delta_1},
 \label{eq:uni3}
\end{eqnarray}
where
\begin{eqnarray}
\delta_1& = &\left(2 \left(g_1(\lambda_e) \lambda _e^2 \left(f_2(\lambda_i)+\alpha  \mu  \left(-2 f_2(\lambda_i)+f_1(\lambda_i) \lambda _i\right) \right)+g_2(\lambda_e) \lambda _e \left(f_1(\lambda_i) \mu  \lambda _i \left(1+2 \alpha  \right)\right.\right.\right. \nonumber\\ && \left.\mbox{} -2 f_2(\lambda_i) \left(-1+\mu +2 \alpha  \mu  \right)\right)+3 g_1(\lambda_e) \left(-f_1(\lambda_i) \mu  \lambda _i \left(1+2 \alpha  \right)\right.\nonumber\\ && \left.\left.\left.\mbox{}+2 f_2(\lambda_i) \left(-1+\mu +2 \alpha  \mu  \right)\right)\right)\right),
\end{eqnarray}
and
\begin{eqnarray}
E_{11}^2& = &\frac{2 E_{11}Ma}{(2+\lambda_e^2Pr_s)}\left(f_2(\lambda_i) \left(-3 g_1(\lambda_e)+g_2(\lambda_e) \lambda _e\right)\right)/\left(g_1(\lambda_e) \lambda _e^2 \left(-f_2(\lambda_i)+\alpha  \mu  \left(2 f_2(\lambda_i)-f_1(\lambda_i) \lambda _i\right) \right)\right. \nonumber\\ && \mbox{}
+3 g_1(\lambda_e) \left(f_1(\lambda_i) \mu  \lambda _i \left(1+2 \alpha \right)-2 f_2(\lambda_i) \left(-1+\mu +2 \alpha  \mu  \right)\right)
\nonumber\\ && \left.\mbox{}
+g_2(\lambda_e) \lambda _e \left(-f_1(\lambda_i) \mu  \lambda _i \left(1+2 \alpha  \right)+2 f_2(\lambda_i) \left(-1+\mu +2 \alpha  \mu  \right)\right)\right).
 \label{eq:uni4}
\end{eqnarray}
The migration velocity is given by
\begin{eqnarray}
\textbf{U}=\textbf{U}_0+Pe_s \textbf{U}_1+Pe_s^2 \textbf{U}_2+O(Pe_s^3).
\label{eq:uni}
\end{eqnarray}
In this case the zeroth order migration velocity $\textsl{\textbf{U}}_0$, given in Eq. (\ref{eq:m1}) reduces to
\begin{eqnarray}
\textbf{U}_0=\frac{3}{2\rho_i+\rho_e}\left[\frac{Y+\mu X+\alpha P}{W+\mu
Z+\alpha G}\right]\left(\frac{3}{2\rho_i+\rho_e}\frac{Y+\mu
X+\alpha P}{W+\mu Z+\alpha G}+\lambda_e^2\right)^{-1}[\vec{{v}}_{0\infty}]_0, \nonumber \\
 \label{eq:m1uni}
\end{eqnarray}
where $\vec{{v}}_{0\infty}$ can be obtained from the relation
\begin{eqnarray}
[\vec{{u}}_{0\infty}]_0 & = & [\vec{{v}}_{0\infty}]_0-\textbf{U}_0\nonumber\\
  & = &\left(1-\frac{3}{2\rho_i+\rho_e}\left[\frac{Y+\mu X+\alpha P}{W+\mu
Z+\alpha G}\right]\left(\frac{3}{2\rho_i+\rho_e}\frac{Y+\mu
X+\alpha P}{W+\mu Z+\alpha G}+\lambda_e^2\right)^{-1}\right)\nonumber\\ && \times [\vec{{v}}_{0\infty}]_0,
 \label{eq:muni1}
\end{eqnarray}
which implies,
\begin{eqnarray}
[\vec{{v}}_{0\infty}]_0&=&\left[1-\frac{3}{2\rho_i+\rho_e}\left[\frac{Y+\mu X+\alpha P}{W+\mu
Z+\alpha G}\right]\left(\frac{3}{2\rho_i+\rho_e}\frac{Y+\mu
X+\alpha P}{W+\mu Z+\alpha G}+\lambda_e^2\right)^{-1}\right]^{-1}\nonumber\\ && \mbox{}e^{\lambda_e^2 t} \hat{i}.
 \label{eq:muni2}
\end{eqnarray}
The first order migration velocity ${\textbf{U}}_1$, given in (\ref{eq:m1f}) reduces to
\begin{eqnarray}
\textbf{U}_1=\frac{3}{2\rho_i+\rho_e}\left[\frac{2 Ma \lambda_e^2f_2(\lambda_i)g_1(\lambda_e)}{(W+\mu Z+\alpha G)}\right]\left(\frac{3}{2\rho_i+\rho_e}\frac{Y+\mu
X+\alpha P}{W+\mu Z+\alpha G}+\lambda_e^2\right)^{-1}E_{11}^1e^{\lambda_e^2 t}\hat{i}, \nonumber
 \label{eq:m1unia}
\end{eqnarray}
and the second order migration velocity $\textsl{\textbf{U}}_2$, given in (\ref{eq:m1s}) reduces to
\begin{eqnarray}
\textbf{U}_2=\frac{3}{2\rho_i+\rho_e}\left[\frac{2 Ma \lambda_e^2f_2(\lambda_i)g_1(\lambda_e)}{(W+\mu Z+\alpha G)}\right]\left(\frac{3}{2\rho_i+\rho_e}\frac{Y+\mu
X+\alpha P}{W+\mu Z+\alpha G}+\lambda_e^2\right)^{-1}E_{11}^2e^{\lambda_e^2 t}\hat{i}. \nonumber
 \label{eq:m1uniab}
\end{eqnarray}

It may be noted that the corresponding migration velocity is only
along the flow direction and avoids cross migration.

We show the
variation of first and second order surfactant distributions with
time (figures \ref{fig:1e} and \ref{fig:1f}). Here, we have
noticed that, the surfactant concentration decreases with time.

For a fixed viscosity ratio, the slip parameter reduces the
resistance offered by the drop. Accordingly, the migration
velocity increases same is observed in figure (\ref{fig:8}). It
may be noted that surface P\'{e}clet number measures the
importance of convection relative to diffusion. Therefore, as
$Pe_s$ increases, migration velocity increases. The same is
observed in figure (\ref{fig:8}). As the viscosity ratio is
increasing, the drop behaves like a solid and hence, the migration
velocity decreases.

%But, for small viscosity ratios as Marangoni number increases, the surface forces dominates the viscous forces. As a consequence, with increasing Marangoni number, the migration velocity decreases. The same is observed in figure (\ref{fig:11}).

It may be noted that Marangoni number is the ratio of surface
tension forces to viscous forces. Therefore, for small viscosity
ratios, with increasing $Ma$, the surface forces dominate and hence,
the drag force increases with the increase of Marangoni number.
Accordingly, the migration velocity decreases with Marangoni
number. But, for large viscosity ratios, the viscous forces
dominates the surface forces. And hence, migration velocity
increases with increasing Marangoni number. The same is observed
in figure (\ref{fig:11}).

We have observed in figures (\ref{fig:1e}) and (\ref{fig:1f}) as time increases, the both first and second order surfactant concentration decreases as expected.

\subsection{Couette flow}\label{cousec}
\begin{figure}
       \begin{subfigure}[b]{0.6\textwidth}
        \centering
{\includegraphics[width=7.2cm, height=7.2cm]{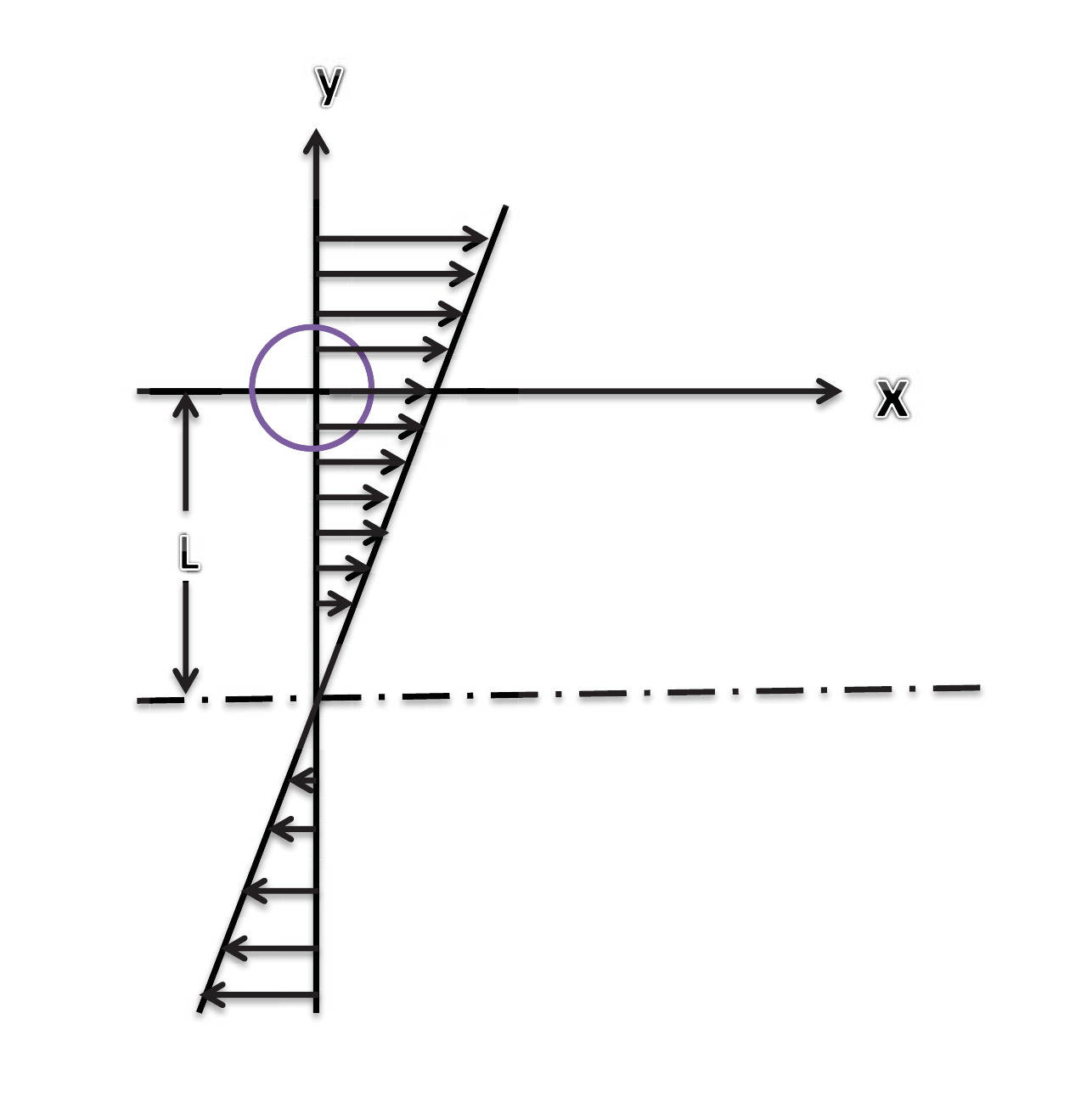}}
%                \caption{A gull}
                %\label{fig:gull}
        \end{subfigure}%
\begin{subfigure}[b]{0.6\textwidth}
               \centering
{\includegraphics[width=7.2cm, height=7.2cm]{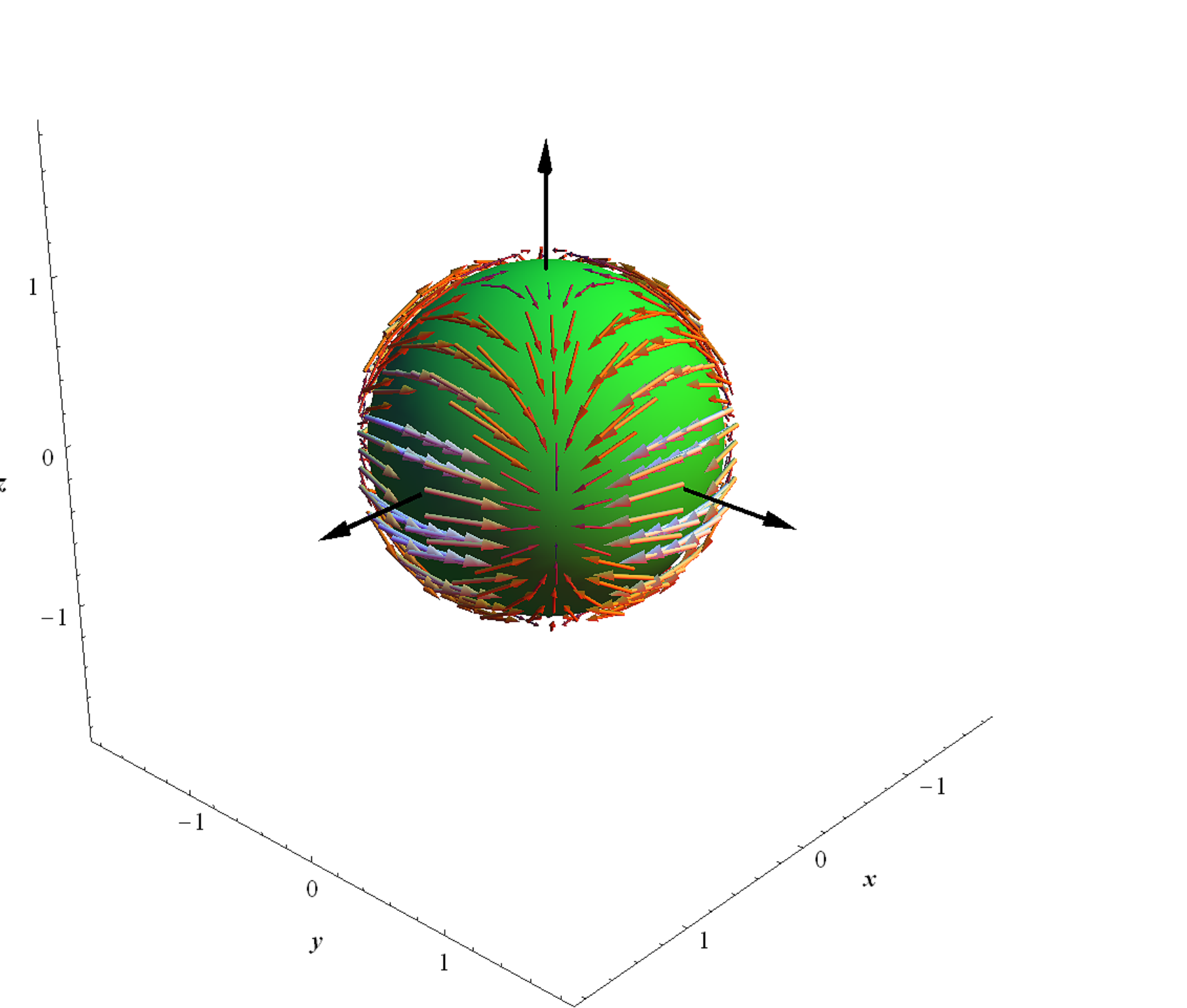}}
%                \caption{A gull2}
               %\label{fig:gull2}
        \end{subfigure}%
         \caption{a) Geometry of the problem and velocity vector corresponding to Coutte ambient flow, b) surface velocity vector field corresponding to Coutte flow}
         \label{fig:coutteflow}
\end{figure}

Consider a Couette flow past a liquid drop of unit radius whose center is at the origin (see Fig. \ref{fig:coutteflow}). In this case, $\vec{{v}}_{\infty}=(F(y+L)\hat{i})e^{\lambda_e^2 t}$, where $L$ is the distance of the center of the
droplet from the point of zero velocity and
$F$ is the shear (Ref. \cite{hetsroni1970flow}).
\begin{figure}
  \centerline{\includegraphics[width=12cm, height=8cm]{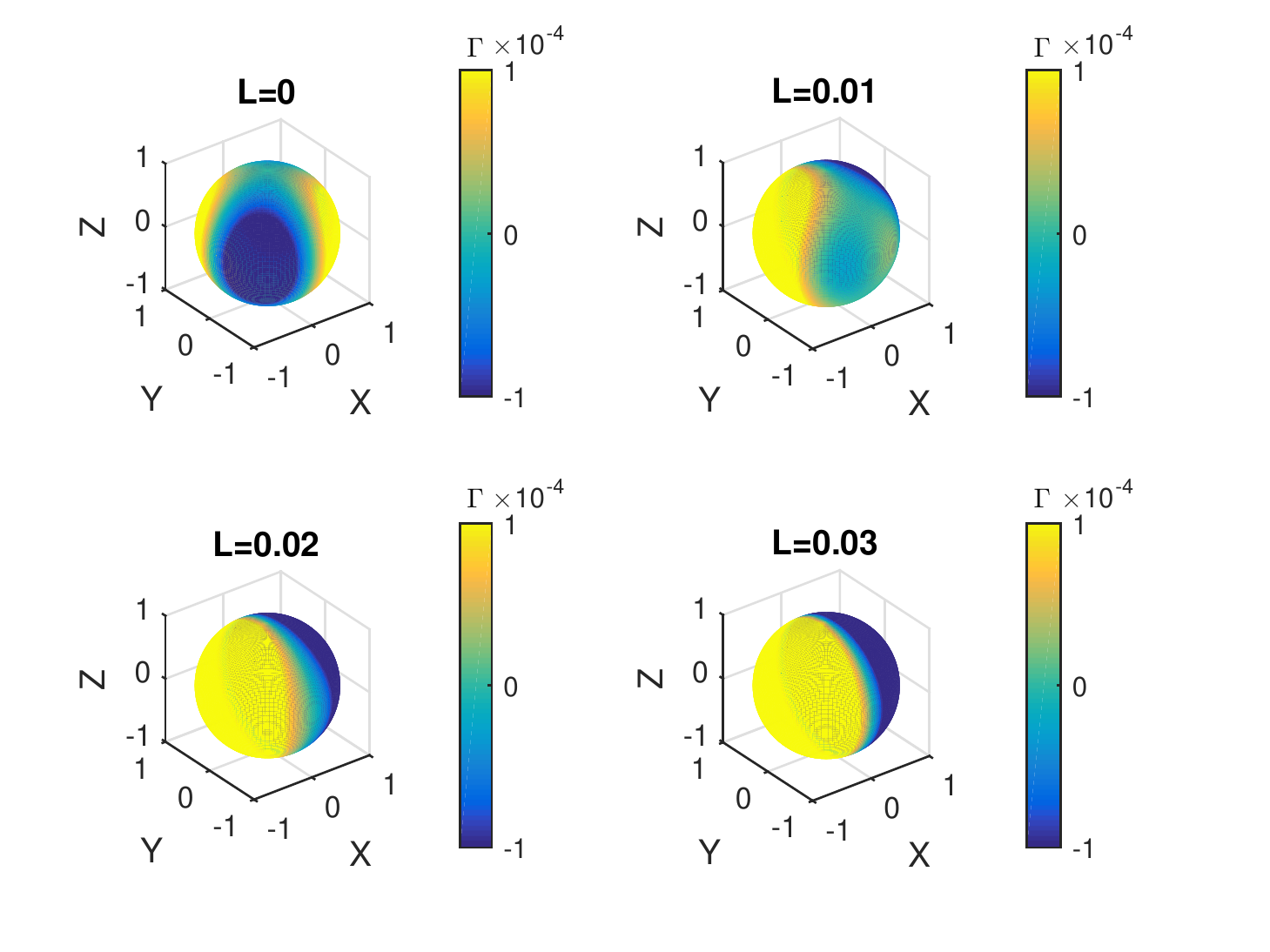}}
  \caption{Variation of first order surfactant distribution for different time values corresponding to Coutte flow, with $\lambda_e^2=-0.04$, $\lambda_i^2=-0.04$, $\mu=5$,$Ma=400$, $F=1$, $t=1$, $Pe=0.01$ and $\alpha=0.1$.}
\label{fig:NCouConL}
\end{figure}

\begin{figure}
  \centerline{\includegraphics[width=12cm, height=8cm]{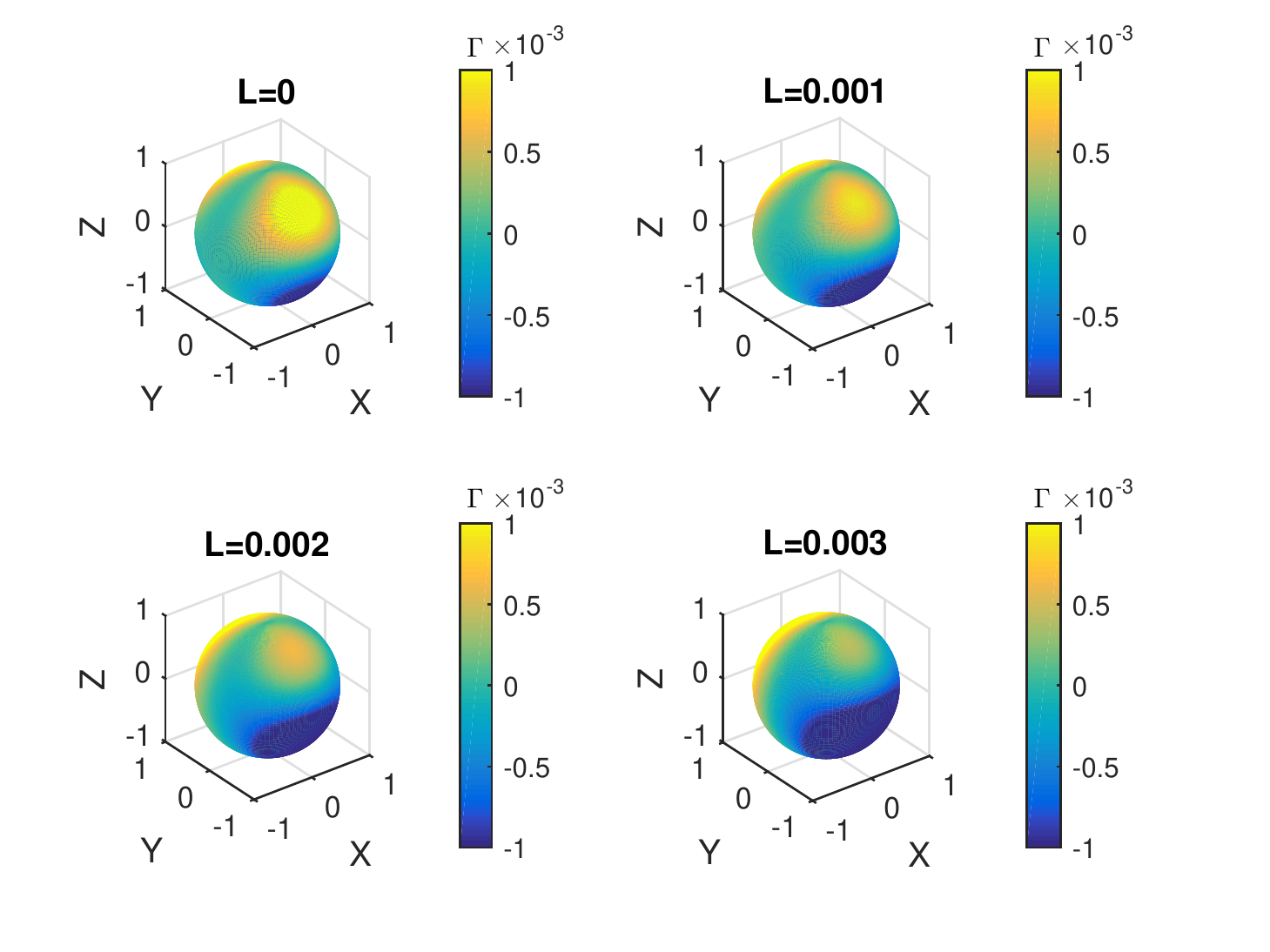}}
  \caption{Variation of second order surfactant distribution for different time values corresponding to Coutte flow, with $\lambda_e^2=-0.04$, $\lambda_i^2=-0.04$, $\mu=5$,$Ma=400$, $F=1$, $t=1$, $Pe=0.01$ and $\alpha=0.1$.}
\label{fig:NCouConL2}
\end{figure}
%\begin{figure}
%  \centerline{\includegraphics[width=12cm, height=8cm]{NConTime1}}
%  \caption{Variation of first order surfactant distribution for different time values corresponding to Coutte flow, with $\lambda_e^2=-0.04$, $\lambda_i^2=-0.04$, $\mu=5$,$Ma=400$, $F=1$, $L=0.1$, $Pe=0.01$ and $\alpha=0.1$.}
%\label{fig:NConTime1}
%\end{figure}
%
%\begin{figure}
%  \centerline{\includegraphics[width=12cm, height=8cm]{NCouConT2}}
%  \caption{Variation of second order surfactant distribution for different time values corresponding to Coutte flow, with $\lambda_e^2=-0.04$, $\lambda_i^2=-0.04$, $\mu=5$,$Ma=400$, $F=1$, $L=0.1$, $Pe=0.01$ and $\alpha=0.1$.}
%\label{fig:NCouConT2}
%\end{figure}
\begin{figure}
  \centerline{\includegraphics[width=7.2cm, height=7.2cm]{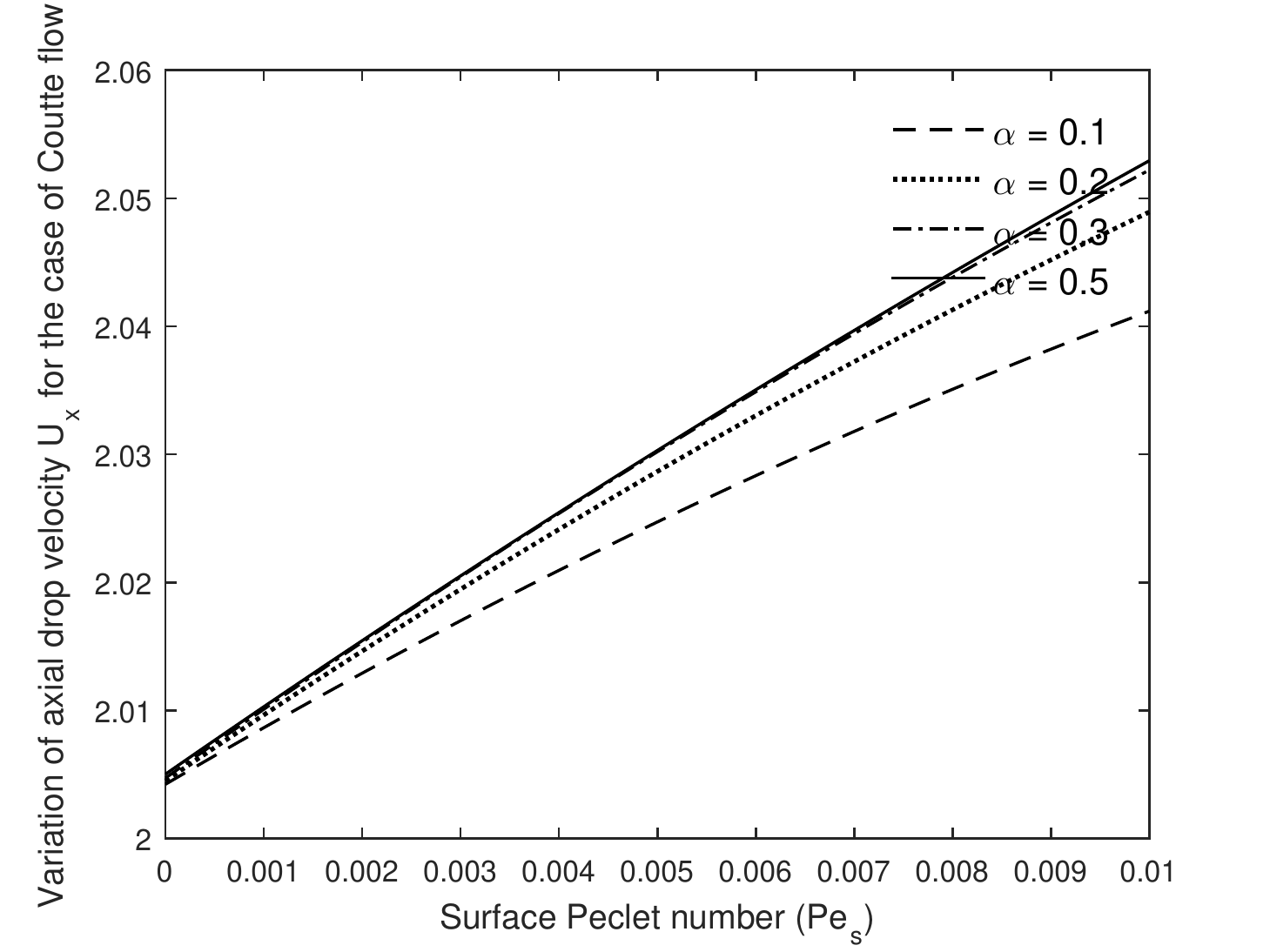}}% Images in 100% size
  \caption{Variation of migration velocity with $Pe_s$ for different slip parameters corresponding to Coutte flow, $\alpha$,  $\lambda_e^2=-0.01$, $\lambda_i^2=-0.01$, $Ma=400$, $F=1$, $L=2$ and $\mu=5$.}
\label{fig:c1}
\end{figure}
\begin{figure}
  \centerline{\includegraphics[width=7.2cm, height=7.2cm]{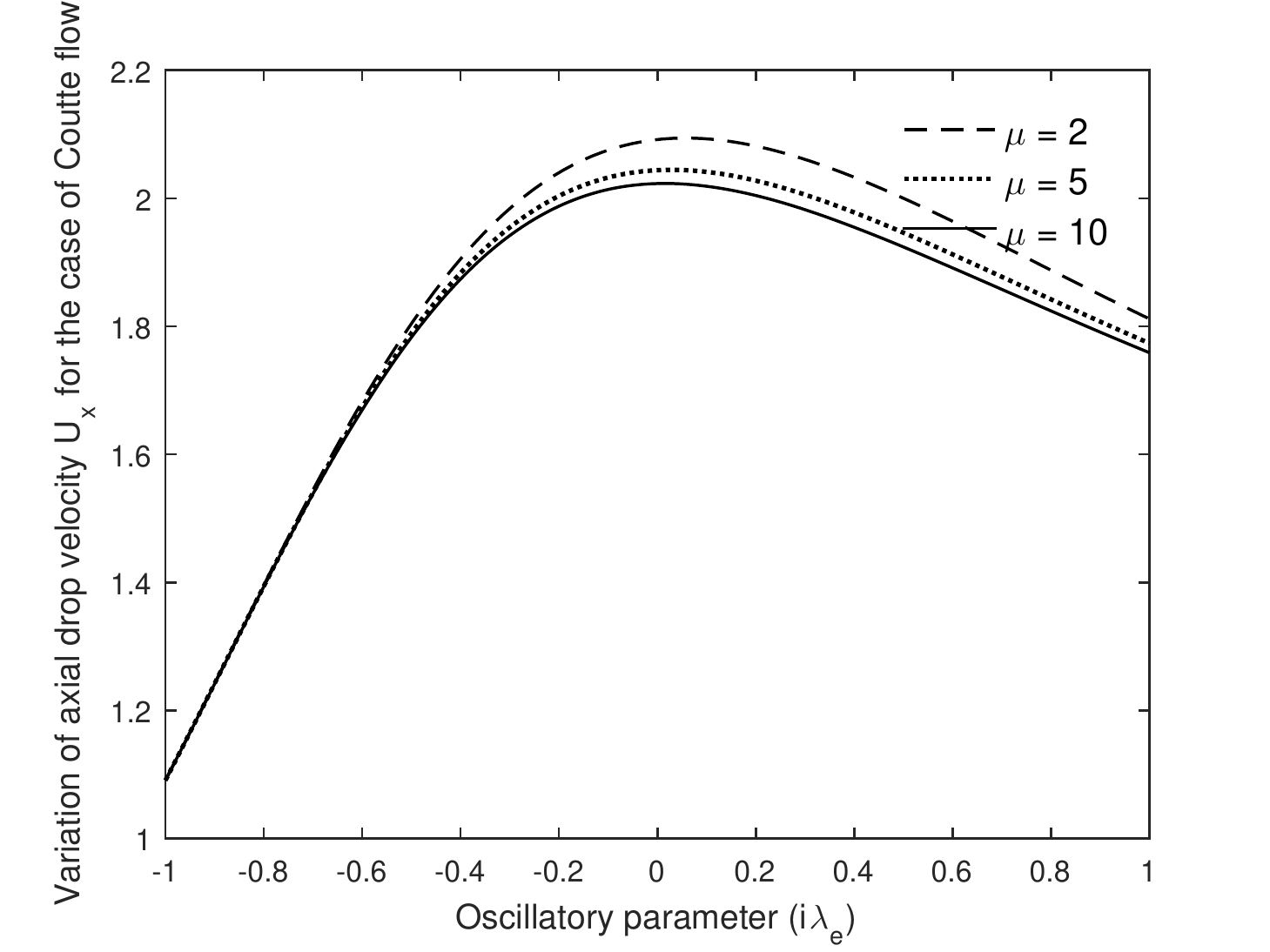}}% Images in 100% size
  \caption{Variation of migration velocity with $i\lambda_e$ for different viscosity ratios corresponding to Coutte flow, $\alpha$, $F=1$, $L=2$, $\lambda_i^2=-0.04$, $Ma=400$ and $\mu=5$.}
\label{fig:c2}
\end{figure}
%\begin{figure}
%  \centerline{\includegraphics[width=7.2cm, height=7.2cm]{NCouLiMu}}% Images in 100% size
%  \caption{Variation of migration velocity with $i\lambda_i$ for different viscosity ratios corresponding to Coutte flow, $\alpha$, $F=1$, $L=2$,  $\lambda_i^2=-0.04$, $Ma=400$ and $\mu=5$.}
%\label{fig:c3}
%\end{figure}

\begin{figure}
       \centerline{\includegraphics[width=7.2cm, height=7.2cm]{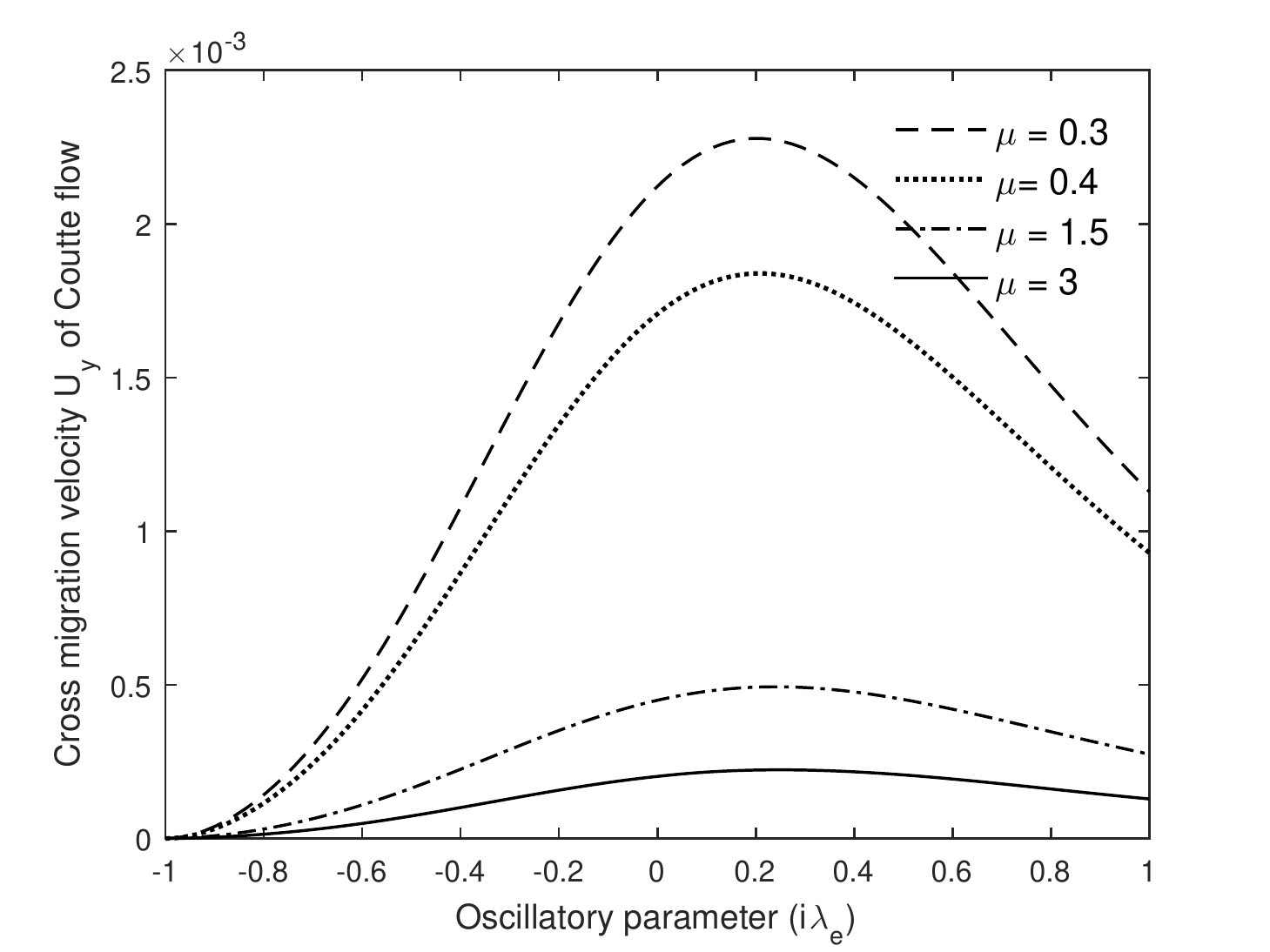}}% Images in 100% size
         \caption{Variation of cross migration velocity with $i\lambda_e$ for different viscosity ratios corresponding to Coutte flow, $\mu$, with $\lambda_i^2=-0.04$, $\alpha=0.2$, $Ma=400$, $F=1$, $L=2$ and $Pe=0.01$.}
         \label{fig:c4}
\end{figure}

%
%\begin{figure}
%       \centerline{\includegraphics[width=7.2cm, height=7.2cm]{NCoucrossLiAl}}% Images in 100% size
%         \caption{Variation of cross migration velocity with $i\lambda_i$ for different slip parameters corresponding to Coutte flow, $\alpha$, with  $i\lambda_e^2=-0.01$, $Ma=400$ and $Pe=0.01$.}
%         \label{fig:c5}
%\end{figure}

\begin{figure}
       \centerline{\includegraphics[width=7.2cm, height=7.2cm]{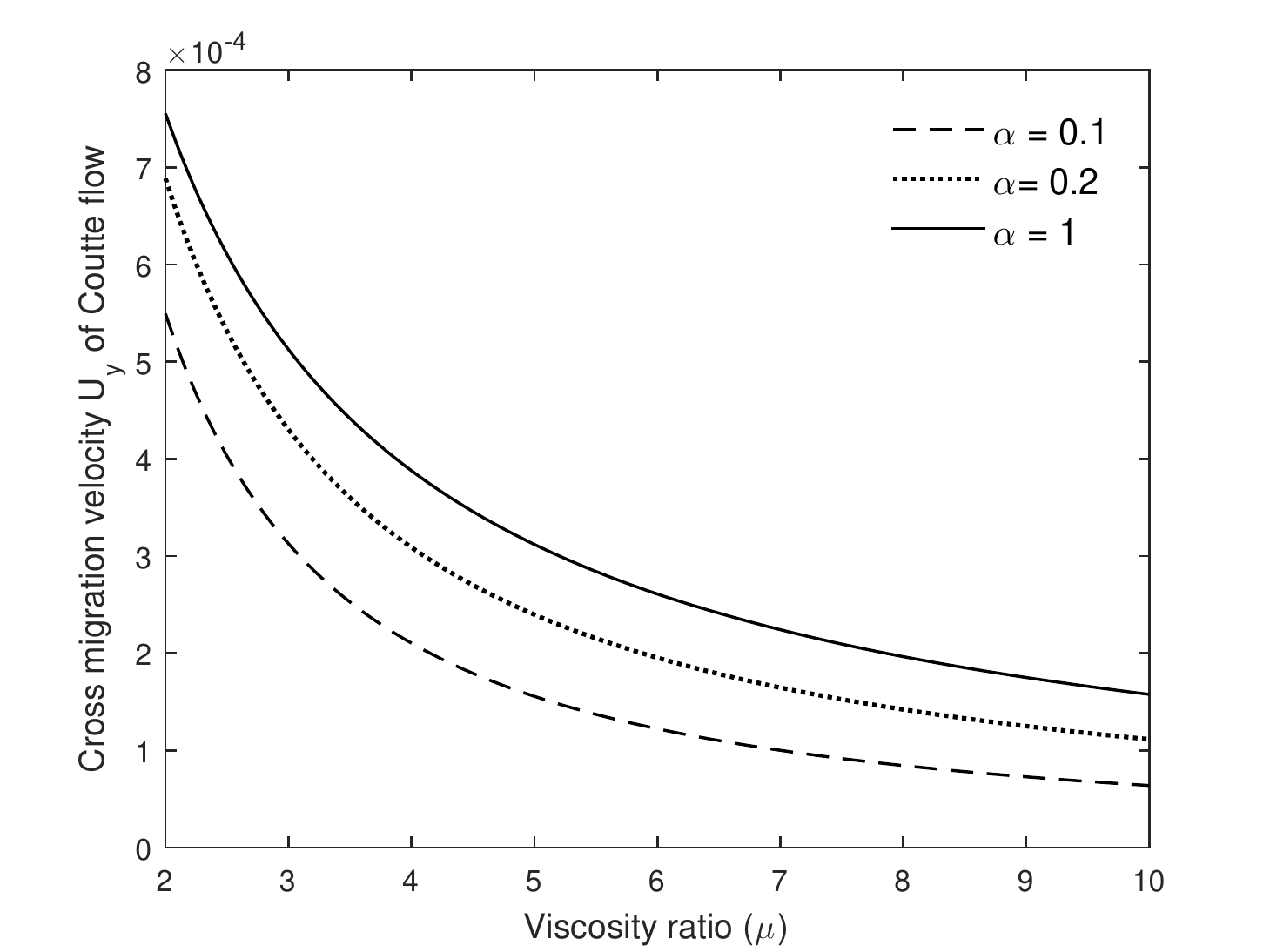}}% Images in 100% size
         \caption{Variation of cross migration velocity with viscosity ratio, $\mu$, for different $\alpha$ corresponding to Coutte flow with $\lambda_e^2=-0.04$, $\lambda_i^2=-0.04$, $\alpha=0.1$, $Ma=400$, $F=1$, $L=4$ and $Pe=0.01$.}
         \label{fig:c6}
\end{figure}
\begin{figure}
  \centerline{\includegraphics[width=7.2cm, height=7.2cm]{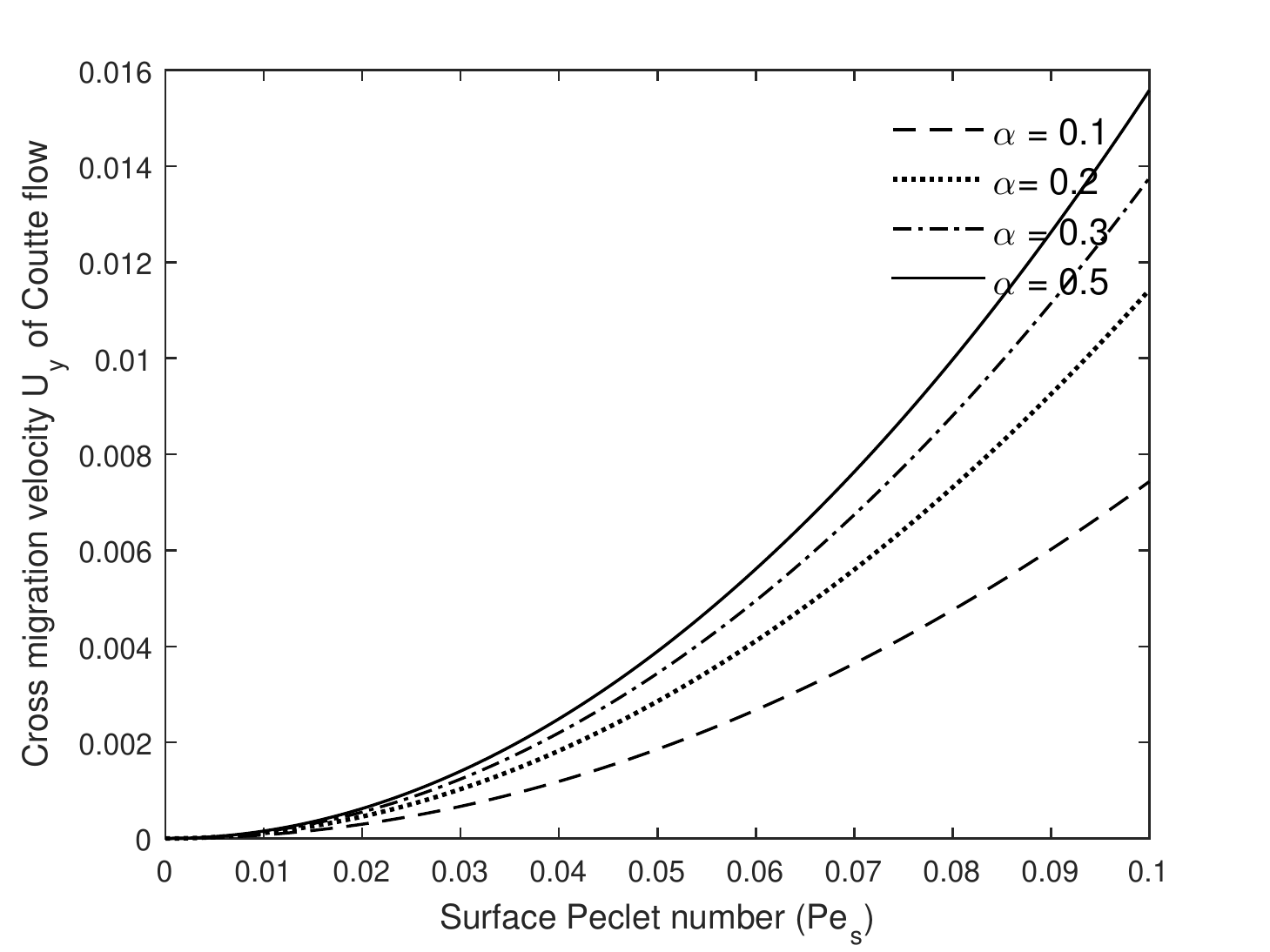}}
  \caption{Variation of cross migration velocity with $Pe_s$ for different slip parameters corresponding to Coutte flow, $\alpha$, with $\lambda_e^2=-0.04$, $\lambda_i^2=-0.04$, $Ma=400$, $F=1$, $L=2$ and $\mu=5$.}
\label{fig:c7}
\end{figure}
\begin{figure}
  \centerline{\includegraphics[width=7.2cm, height=7.2cm]{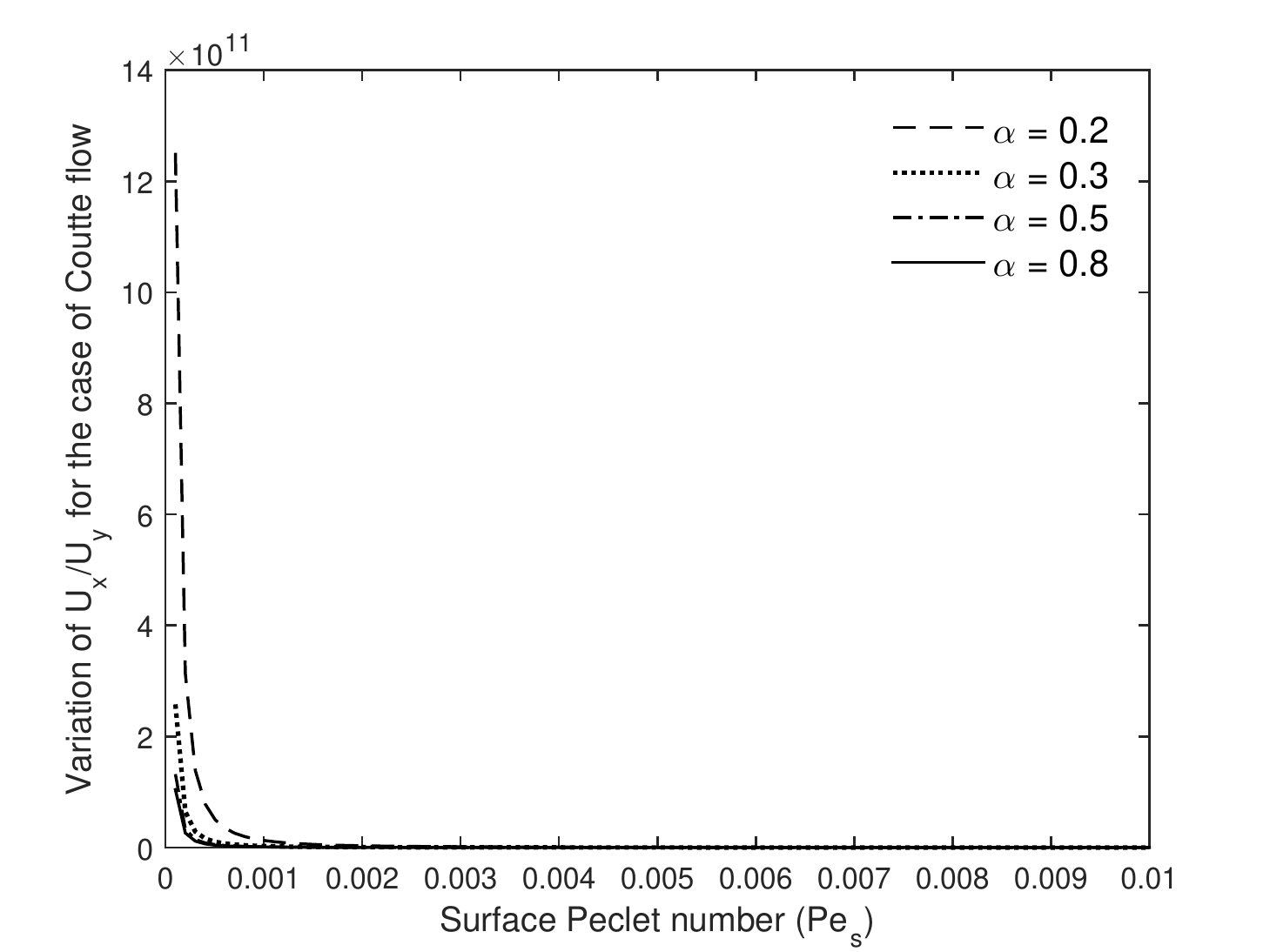}}
  \caption{Variation of $U_x/U_y$ with $Pe_s$ for different slip parameters corresponding to Coutte flow, $\alpha$, with $\lambda_e^2=-0.04$, $\lambda_i^2=-0.04$, $Ma=400$, $F=1$, $L=2$ and $\mu=5$.}
\label{fig:cuxuy}
\end{figure}

 Therefore the corresponding scalar functions $\chi_0^\infty$ and $\eta_0^\infty$ are given by
\begin{eqnarray}
\chi_0^\infty=\left(\frac{FL}{2}r P_1^1(\cos\,\theta) \cos\,\phi+\frac{F}{36}r^2 \sin\, 2 \phi P_2^2(\cos\,\theta)\right) e^{\lambda_e^2 t},\,\,\,\,\,\eta_0=0.\nonumber
\end{eqnarray}
The above choice indicates that $\alpha_{1}^0=\frac{FL}{2}$, $\alpha_{2}^0=\frac{F}{36}$,
$\beta_{1}^0=0$, $\gamma_{1}^0=0$  in Eqs.~(\ref{eq:19}) and
(\ref{eq:20}). Therefore
the corresponding surfactant concentration distribution on the
spherical drop is given by
\begin{eqnarray}
\Gamma=\Gamma_0+Pe_s \Gamma_1+Pe_s^2 \Gamma_2+O(Pe_s^3),
\label{eq:conCout}
\end{eqnarray}
where
\begin{eqnarray}
\Gamma_0=1,
\label{eq:conCout1}
\end{eqnarray}
\begin{eqnarray}
\Gamma_1=\left(-E_{11}^1\sin\,\theta \cos\,\phi+F_{22}^1 \sin\, 2 \phi P_2^2(\cos\,\theta)\right) e^{\lambda_e^2 t},
\label{eq:conCout1}
\end{eqnarray}
\begin{eqnarray}
\Gamma_2&=&\left(-E_{11}^2\sin\,\theta \cos\,\phi-F_{11}^2 \sin\,\theta \sin\,\phi+F_{22}^2 \sin\, 2 \phi P_2^2(\cos\,\theta)\right.
\nonumber\\  &&
\mbox{}
+F_{31}^2 \sin\,\phi P_3^1(\cos\,\theta)
+ F_{33}^2  \sin\, 3 \phi P_3^3(\cos\,\theta)+E_{20}^2 P_2^0(\cos\,\theta)\nonumber\\  &&\left.
\mbox{}+E_{22}^2 \cos\, 2 \phi P_2^2(\cos\,\theta)+E_{40}^2 P_4^0(\cos\,\theta)+E_{44}^2  \cos\, 4 \phi P_4^4(\cos\,\theta)\right)\nonumber\\  &&
\mbox{} e^{\lambda_e^2 t},
\label{eq:conCout1}
\end{eqnarray}
where few quantities $E_{11}^1, F_{22}^1$ etc. are listed in the
Appendix (\ref{appD}).

For an unbounded Coutte flow, the migration velocity of a force free drop is calculated as
\begin{eqnarray}
\textbf{U}=\textbf{U}_0+Pe_s \textbf{U}_1+Pe_s^2 \textbf{U}_2+O(Pe_s^3).
\label{eq:uni}
\end{eqnarray}
In this case the zeroth order migration velocity ${\textbf{U}}_0$, given in Eq. (\ref{eq:m1}) reduces to
\begin{eqnarray}
\textbf{U}_0=\frac{3}{2\rho_i+\rho_e}\left[\frac{Y+\mu X+\alpha P}{W+\mu
Z+\alpha G}\right]\left(\frac{3}{2\rho_i+\rho_e}\frac{Y+\mu
X+\alpha P}{W+\mu Z+\alpha G}+\lambda_e^2\right)^{-1}FLe^{\lambda_e^2 t}\hat{i}, \nonumber\\
 \label{eq:m1couette}
\end{eqnarray}

The first order migration velocity ${\textbf{U}}_1$, given in (\ref{eq:m1f}) reduces to
\begin{eqnarray}
{\textbf{U}}_1=\frac{3E_{11}^1\hat{i}}{2\rho_i+\rho_e}\left[\frac{2 Ma \lambda_e^2f_2(\lambda_i)g_1(\lambda_e)}{(W+\mu Z+\alpha G)}\right]\left(\frac{3}{2\rho_i+\rho_e}\frac{Y+\mu
X+\alpha P}{W+\mu Z+\alpha G}+\lambda_e^2\right)^{-1}e^{\lambda_e^2 t}, \nonumber\\
 \label{eq:m1couettea}
\end{eqnarray}
and the second order migration velocity ${\textbf{U}}_2$, given in (\ref{eq:m1s}) reduces to
\begin{eqnarray}
{\textbf{U}}_2=\frac{3(E_{11}^2\hat{i}+F_{11}^2\hat{j})}{2\rho_i+\rho_e}\left[\frac{2 Ma \lambda_e^2f_2(\lambda_i)g_1(\lambda_e)}{(W+\mu Z+\alpha G)}\right]\left(\frac{3}{2\rho_i+\rho_e}\frac{Y+\mu
X+\alpha P}{W+\mu Z+\alpha G}+\lambda_e^2\right)^{-1}e^{\lambda_e^2 t}. \nonumber\\
 \label{eq:m1couetteb}
\end{eqnarray}

By symmetry, it is expected that there will be no velocity
component in $z$ direction. In case if there is a cross migration
(i.e., motion transverse to the flow direction), the same occurs
towards the center line and should be in $y$ direction (see
Fig. (\ref{fig:coutteflow})). In \cite{pak2014viscous}, a detailed explanation on cross
migration of a surfactant coated viscous drop in Poiseuille flow
is presented. Similar arguments followed in \cite{mandal2015effect} while
discussing migration of deformed drop with interfacial slip in in an unbounded Poiseuille flow.  We also follow similar arguments to show that cross
migration occurs only at second order with respect to the
expansion of migration velocity in terms of surface P\'{e}clet
number. Here, we have observed that, at leading order, we recover
the case of clean spherical drop in an unbounded Couette flow
(characterized by the velocity scale $U_c$). It is well known
that, there can be no cross stream migration in the absence of
inertia, deformation, and surfactant concentration. Accordingly,
we observe that at leading order, there is no cross stream
migration. This phenomena can also be supported mathematically as
follows: The dimensional zeroth order migration velocity
${\textbf{U}}^*_0={\textbf{U}}_0U_c\propto U_c$. If there is a
cross stream migration, for a reversal of the background flow
direction ($U_c\rightarrow -U_c$), ${\textbf{U}}^*_0$ also should
change the direction. But, the cross stream migration should occur
towards the center line. Therefore, there is no cross stream
migration in leading order, which implies the symmetry condition
at leading order is satisfied (Ref. \cite{pak2014viscous}).

With the similar argument which is given for leading order
migration velocity, we can say at first order also there can not
be cross stream migration. We observe the dimensional first order
migration velocity ${\textbf{U}}^*_1=Pe_s{\textbf{U}}_1 U_c
\propto Pe_s Ma E_{11}^1 U_c\propto Pe_s Ma U_c$, and the product
$Pe_s Ma$ is independent of $U_c$. Therefore,
${\textbf{U}}^*_1\propto U_c$.  If there is a cross migration
velocity, ${\textbf{U}}^*_1.\hat{j}\propto U_c$, which violates
the symmetry requirement that the cross stream migration direction
remains same upon reversal of the background flow direction.
Lateral migration is therefore expected not to occur at leading
and first order.

Observing second order migration velocity, we see
${\textbf{U}}^*_2.\hat{i}=Pe_s^2{\textbf{U}}_2 U_c.\hat{i}\propto
Pe_s^2 Ma E_{11}^2 U_c\propto Pe_s^2 Ma^2 U_c\propto U_c$, and
${\textbf{U}}^*_2.\hat{j}=Pe_s^2{\textbf{U}}_2 U_c.\hat{j}\propto
Pe_s^2 Ma F_{11}^2 U_c\propto Pe_s^2 Ma U_c\propto U_c^2$.
Therefore, the transverse migration is invariant upon reversal of
the direction of the ambient Coutte flow.

We also noted that, the transverse migration is linearly dependent
on $L$ (as $E_{11}^1\propto L$) which respects the symmetry
requirement that the transverse migration direction should reverse
its sign when the drop is placed at the same distance but on the
opposite side with respect to the center of the Coutte flow (see
Fig. \ref{fig:coutteflow}). The same is noted by Pak et al.
\cite{pak2014viscous} for the case of Poiseuille flow.

The first order and second order surfactant distributions are plotted with specific values of parameters for visualization in figures (\ref{fig:NCouConL}) and (\ref{fig:NCouConL2}). Here, we have seen as $L$ increases, the concentration increases (since $E_{11}^1\propto L$). 

We have observed the variation of axial migration velocity and
cross-stream migration velocity for different parameters. The
variation of axial migration is observed in figures
(\ref{fig:c1}) and (\ref{fig:c2}). It can be seen
that, we have the cross migration due to the linear term present
in the ambient velocity of Coutte flow. And the migration in the
axial direction is due to the constant term present in the Coutte
flow. As a consequence, axial migration velocity in the case of
Coutte flow behaves in the same manner as the migration velocity
in the case uniform ambient flow.

We have observed the variation of cross
migration velocity with the amplification factor $\lambda_e$ in figure (\ref{fig:c4}). Drop
cross migration is oscillating with the
amplification factor. From Fig. (\ref{fig:c6}), we have seen
that, with the increasing viscosity ratio, the magnitude of cross
migration decreases as expected (as the viscosity ratio increases,
the drop behaves like a solid).

It may be noted that surface P\'{e}clet number measures the
importance of convection relative to diffusion. Therefore, as
$Pe_s$ increases, magnitude of migration velocity increases. Also,
for a fixed viscosity ratio, the slip parameter reduces the
resistance offered by the drop. Accordingly, the migration
velocity increases same is observed in figure (\ref{fig:c7}). From figure (\ref{fig:cuxuy}), we have observed the ratio of $U_x$ and $U_y$ decreases. From this, we can say, $U_y$ increases faster than $U_x$ with P\'{e}clet number.

\subsection{Poiseuille flow}
Consider a Poiseuille flow past a liquid drop of unit radius whose center is at origin (Ref. \cite{hetsroni1970flow,pak2014viscous} to see the geometrical setup of the problem). In this case, we calculated the ambient velocity as
$\vec{{v}}_{\infty}=\vec{{v}}_{0\infty}=-\hat{k}e^{\lambda_e^2t}
\left(1-\frac{J_{0}(i\lambda_eR)}{J_{0}(i\lambda_eR_0)}\right)
\left(1-\frac{1}{J_{0}(i\lambda_eR_0)}\right)^{-1}$. Here $R^2=r^2 \sin^2\,\theta+b^2+2br\sin\,\theta\cos\,\phi $, velocity is non-dimensionalized with the characteristic velocity $U_b$, which is at a dimensionless distance $b$ from the drop, and $R_0$ is the dimensionless distance to the point of zero velocity of the flow, $\lambda_e$ is the amplification factor. We expanded $\vec{{v}}_{\infty}$ as series form for small $\lambda_e$ to get $\chi_0^\infty$ and $\eta_0^\infty$, which are given by
\begin{eqnarray}
\chi_0^\infty=\left[\beta_{12}f_1(\lambda_{e}r)S_{1}(\theta,\phi)+\sum\limits_{n=1}^\infty \alpha_{n2}r^{n}S_{n}(\theta,\phi)\right]e^{\lambda_e^2 t},\,\,\,\,\,\eta_0^\infty=0,\nonumber
\end{eqnarray}
where,
\begin{eqnarray}
\alpha_{1}^0=-\frac{1}{2}\left(1-\frac{1}{J_{0}(i\lambda_eR_0)}\right)^{-1}
\left(1-\frac{b^2\lambda_e^2}{4J_{0}(i\lambda_eR_0)}\right),
\end{eqnarray}
\begin{eqnarray}
\alpha_{2}^0=-\left(1-\frac{1}{J_{0}(i\lambda_eR_0)}\right)^{-1}
\left(\frac{b\lambda_e^2}{36J_{0}(i\lambda_eR_0)}\right)
\end{eqnarray}
\begin{eqnarray}
\alpha_{3}^0=-\left(\frac{\lambda_e^2}{120J_{0}(i\lambda_eR_0)}\right)
\left(1-\frac{1}{J_{0}(i\lambda_eR_0)}\right)^{-1}
\end{eqnarray}
\begin{eqnarray}
\beta_{1}^0=\left(\frac{3}{2\lambda_eJ_{0}(i\lambda_eR_0)}\right)
\left(1-\frac{1}{J_{0}(i\lambda_eR_0)}\right)^{-1}
\end{eqnarray}
 Therefore
the corresponding surfactant concentration distribution on the
spherical drop is given by
\begin{eqnarray}
\Gamma=\Gamma_0+Pe_s \Gamma_1+Pe_s^2 \Gamma_2+O(Pe_s^3),
\label{eq:conPoi}
\end{eqnarray}
where
\begin{eqnarray}
\Gamma_0=1,
\label{eq:conPoi1}
\end{eqnarray}
\begin{eqnarray}
\Gamma_1=\left(E_{10}^1\cos\,\theta+E_{21}^1 \cos\, \phi P_2^1(\cos\,\theta)+E_{30}^1P_3^0(\cos\,\theta)\right) e^{\lambda_e^2 t},
\label{eq:conPois1}
\end{eqnarray}
\begin{eqnarray}
\Gamma_2&=&\left(E_{10}^2\cos\,\theta+E_{21}^2 \cos\, \phi P_2^1(\cos\,\theta)+E_{30}^2P_3^0(\cos\,\theta)\right.
\nonumber\\  &&
\mbox{}
+E_{11}^2P_1^1(\cos\,\theta) \cos\,\phi+E_{20}^2 P_2^0(\cos\,\theta)+E_{22}^2 \cos\, 2 \phi P_2^2(\cos\,\theta)
\nonumber\\  &&
\mbox{}
+E_{31}^2 \cos\,\phi P_3^1(\cos\,\theta)
+ E_{42}^2  \cos\, 2 \phi P_4^2(\cos\,\theta)+E_{51}^2 \cos\,\phi P_5^1(\cos\,\theta)
\nonumber\\  &&
\mbox{}
+E_{62}^2 \cos\,2\phi P_6^2(\cos\,\theta)
+ E_{71}^2  \cos\, \phi P_7^1(\cos\,\theta)+E_{82}^2 \cos\,2\phi P_8^2(\cos\,\theta)
\nonumber\\  &&\left.
\mbox{}+E_{52}^2 \cos\, 2 \phi P_5^2(\cos\,\theta)+E_{40}^2 P_4^0(\cos\,\theta)+E_{60}^2  P_6^0(\cos\,\theta)\right)\nonumber\\  &&
\mbox{} e^{\lambda_e^2 t},
\label{eq:conPois1}
\end{eqnarray}
where the constants $E_{10}^1$, $E_{21}^1$, etc can be computed  Eqs.~(\ref{enm}),
(\ref{fnm}), (\ref{inm}) and
(\ref{jnm}).

For an unbounded Poiseuille flow, the migration velocity of a force free drop is calculated as
\begin{eqnarray}
\textbf{U}=\textbf{U}_0+Pe_s \textbf{U}_1+Pe_s^2 \textbf{U}_2+O(Pe_s^3).
\label{eq:Poi}
\end{eqnarray}
In this case the zeroth order migration velocity ${\textbf{U}}_0$, given in Eq. (\ref{eq:m1}) reduces to
\begin{eqnarray}
\textbf{U}_0 & = & \frac{3}{2\rho_i+\rho_e}\left[\frac{Y+\mu X+\alpha P}{W+\mu
Z+\alpha G}[\vec{{v}}_{0\infty}]_0+\frac{V+\mu U+\alpha H}{W+\mu Z+\alpha G}[\nabla^2
\vec{{v}}_{0\infty}]_0 \right] \nonumber\\
&& \left(\frac{3}{2\rho_i+\rho_e}\frac{Y+\mu
X+\alpha P}{W+\mu Z+\alpha G}+\lambda_e^2\right)^{-1}, \nonumber\\
 \label{eq:m1poi}
\end{eqnarray}
where
\begin{eqnarray}
[\vec{{v}}_{0\infty}]_0& = & \left(\frac{I_0\left(b \lambda_e\right)-J_0(i\lambda_e R_0)}{(1-J_0(\lambda_e R_0)) J_0(i\lambda_e R_0)}\right)e^{\lambda_e^2 t} \hat{k}.
 \label{eq:mpoi2}
\end{eqnarray}
and
\begin{eqnarray}
[\nabla^2
\vec{{v}}_{0\infty}]_0& = & \frac{\lambda_e^2 I_0\left(b \lambda_e\right)}{I_0(\lambda_e R_0)-I_0(\lambda_e R_0)^2}e^{\lambda_e^2 t} \hat{k}.
 \label{eq:mcouette2}
\end{eqnarray}
If we consider the limiting case of no oscillations in the hydrodynamic flow field, i.e., $\lambda_i=\lambda_e=0$, and zero slip coefficient, i.e., $\alpha=0$, then the zeroth order terminal velocity reduces to
\begin{equation}
\textbf{U}_0=\left(1-\frac{b^2}{R_0^2}-\frac{\mu}{4+6\mu}\frac{4}{R_0^2}\right)\hat{k},
\label{eq:m2comp1}
\end{equation}
which is exactly matching with the one that is obtained by Pak, Feng and Stone \cite{pak2014viscous}.

The first order migration velocity ${\textbf{U}}_1$, given in (\ref{eq:m1f}) reduces to
\begin{eqnarray}
{\textbf{U}}_1=\frac{3E_{10}^1\hat{k}}{2\rho_i+\rho_e}\left[\frac{2 Ma \lambda_e^2j_2(\lambda_i)h_1(\lambda_e)}{(W+\mu Z+\alpha G)}\right]\left(\frac{3}{2\rho_i+\rho_e}\frac{Y+\mu
X+\alpha P}{W+\mu Z+\alpha G}+\lambda_e^2\right)^{-1}e^{\lambda_e^2 t}, \nonumber\\
 \label{eq:m1couettea}
\end{eqnarray}
and the second order migration velocity ${\textbf{U}}_2$, given in (\ref{eq:m1s}) reduces to
\begin{eqnarray}
{\textbf{U}}_2=\frac{3(E_{11}^2\hat{i}+E_{10}^2\hat{k})}{2\rho_i+\rho_e}\left[\frac{2 Ma \lambda_e^2j_2(\lambda_i)h_1(\lambda_e)}{(W+\mu Z+\alpha G)}\right]\left(\frac{3}{2\rho_i+\rho_e}\frac{Y+\mu
X+\alpha P}{W+\mu Z+\alpha G}+\lambda_e^2\right)^{-1}e^{\lambda_e^2 t}. \nonumber\\
 \label{eq:m1couetteb}
\end{eqnarray}

Similar to the arguments made in Section (\ref{cousec}),
by symmetry, it is expected that there will be no velocity
component in $y$ direction. In case if there is a cross migration, the same occurs
towards the center line and should be in $x$ direction (Ref.
\cite{pak2014viscous}). We also follow similar arguments to show that, there is no cross stream
migration in leading order and first order, which implies the symmetry condition
at leading order is satisfied (Ref. \cite{pak2014viscous}).

Similarly, observing second order migration velocity, we see
${\textbf{U}}^*_2.\hat{k}=Pe_s^2{\textbf{U}}_2 U_c.\hat{k}\propto
Pe_s^2 Ma E_{10}^2 U_c\propto Pe_s^2 Ma^2 U_c\propto U_c$, and
${\textbf{U}}^*_2.\hat{i}=Pe_s^2{\textbf{U}}_2 U_c.\hat{i}\propto
Pe_s^2 Ma E_{11}^2 U_c\propto Pe_s^2 Ma U_c\propto U_c^2$.
Therefore, the transverse migration is invariant upon reversal of
the direction of the ambient Poiseuille flow.

%\section{Results and discussion}

\subsection{Validation}
We have compared our results with some existing literature to validate our results. These are shown in the Table (\ref{table1}).
\begin{table}[]
\centering \caption{Limiting cases of the magnitude of drag force of
the present study to get that of different existing literature. \vspace{0.25cm}} \label{table1}
\renewcommand{\baselinestretch}{1.50}\normalsize
\begin{tabular}{|p{2cm}|p{4cm}|p{5cm}|}\hline
                     & Main contribution             & Limiting cases of current study to get others as listed    \\ \hline
    Present study    &  Effect of surfactant concentration and interfacial slip $\alpha$ on the unsteady Stokes flow past a viscous drop for low $Pe_s$ (Surface P\'{e}clet numbar).         &                  \\ \hline
    Pak et al. \cite{pak2014viscous}    &  Effect of surfactant concentration on the steady Stokes flow past a viscous drop for low $Pe_s$ (Surface P\'{e}clet numbar).     &  $D(Pe_s,Ma,\mu,\alpha\rightarrow0,Pr_s\rightarrow0,\lambda_e\rightarrow0,\lambda_i\rightarrow0,t)=D_1(Pe_s,Ma,\mu)$                \\ \hline
    Sharanya and Raja Sekhar \cite{sharanya2015thermocapillary}    & Thermocapillary migration of a spherical drop in an arbitrary transient Stokes flow           &  $D(Pe_s\rightarrow0,Ma\rightarrow0,\mu,\alpha\rightarrow0,Pr_s,\lambda_e,\lambda_i,t)=D_2(Ma\rightarrow0,\mu,Pr,\lambda_e,\lambda_i,t)$                  \\ \hline
    Choudhuri and Raja Sekhar \cite{choudhuri2013thermocapillary}    &   Thermocapillary migration of a spherical drop in an arbitrary transient Stokes flow       &    $D(Pe_s\rightarrow0,Ma\rightarrow0,\mu,\alpha\rightarrow0,Pr_s\rightarrow0,\lambda_e\rightarrow0,\lambda_i\rightarrow0,t)=D_3(Ma\rightarrow0,\mu)$               \\ \hline
     Choudhuri and Padmavati \cite{choudhuri2014study}    &   Oscillatory Stokes flow past a viscous drop    &    $D(Pe_s\rightarrow0,Ma\rightarrow0,\mu,\alpha\rightarrow0,Pr_s\rightarrow0,\lambda_e,\lambda_i,t)=D_4(\lambda_e,\lambda_i,\mu)$               \\ \hline
\end{tabular}
\end{table}

\section{Conclusions}
In this paper, we have considered an arbitrary transient Stokes flow with a given ambient flow past a spherical drop. We analyzed the effects of surface-active agents on the motion of the drop. We have solved the unsteady convection-diffusion equation to find the surfactant transport on the surface of the drop for low surface P\'{e}clet number. We have also considered the effects of interfacial slip. We found a closed form expression for drag and migration velocity in terms of Marangoni number, slip parameter and viscosity ratios up to second order in the surface P\'{e}clet number, i.e., up to $O(Pe_s^2)$. We have analyzed the variation of surfactants for different viscosity ratios and Marangoni number. We have observed that the impurities residing on the surface do not show much effect on the behavior of the drop for increasing viscosity ratios. We considered various special cases and computed drag and migration velocity up to second order in the surface P\'{e}clet number in each case. We have also compared the results with the existing literature for some limiting cases.

\section{acknowledgements}
One of the authors (VS) would like to acknowledge the financial support by CSIR-UGC (F.No. 17-06/2012 (i) EU-V dated, 05-10-2012), India.

\appendix
\section{The unknown coefficients (for leading order problem)}\label{appA}
The unknown coefficients in (\ref{eq:21}) and (\ref{eq:24}) can be found using the boundary conditions given in (\ref{eq:14}) to (\ref{eq:18}) which are given as follows:
\begin{eqnarray}
\hat{\alpha}_{n}^0& = &\left(-2 f_{n+1}(\lambda_i) g_{n+1}(\lambda_e) \alpha _{n}^0 \lambda _e+2 f_{n+1}(\lambda_i) g_2(\lambda_e) \mu  \alpha _{n}^0 \lambda _e-2 f_{n+1}(\lambda_e) f_{n+1}(\lambda_i) g_n(\lambda_e) \beta _{n}^0 \lambda _e\right. \nonumber\\
  &&  \mbox{}
-2 f_n(\lambda_e) f_{n+1}(\lambda_i) g_{n+1}(\lambda_e) \beta _{n}^0 \lambda _e+2 f_{n+1}(\lambda_e) f_{n+1}(\lambda_i) g_n(\lambda_e) \mu  \beta _{n}^0 \lambda _e+2 f_n(\lambda_e) f_{n+1}(\lambda_i) g_{n+1}(\lambda_e) \mu  \beta _{n}^0 \lambda _e\nonumber\\
  &&  \mbox{}
-f_{n+1}(\lambda_i) g_n(\lambda_e) \alpha _{n}^0 \lambda _e^2-f_n(\lambda_i) g_{n+1}(\lambda_e) \mu  \alpha _{n}^0 \lambda _e \lambda _i-f_n(\lambda_i) f_{n+1}(\lambda_e) g_n(\lambda_e) \mu  \beta _{n}^0 \lambda _e \lambda _i\nonumber\\
  &&  \mbox{}
-f_n(\lambda_e) f_n(\lambda_i) g_{n+1}(\lambda_e) \mu  \beta _{n}^0 \lambda _e \lambda _i+4 f_{n+1}(\lambda_i) g_{n+1}(\lambda_e) \alpha  \mu  \alpha _{n}^0 \lambda _e+4 f_{n+1}(\lambda_e) f_{n+1}(\lambda_i) g_n(\lambda_e) \alpha  \mu  \beta _{n}^0 \lambda _e
\nonumber\\
  &&  \mbox{}
+4 f_n(\lambda_e) f_{n+1}(\lambda_i) g_{n+1}(\lambda_e) \alpha  \mu  \beta _{n}^0 \lambda _e+2 f_{n+1}(\lambda_i) g_n(\lambda_e) \alpha  \mu  \alpha _{n}^0 \lambda _e^2-2 f_n(\lambda_i) g_{n+1}(\lambda_e) \alpha  \mu  \alpha _{n}^0 \lambda _e \lambda _i
\nonumber\\
  &&  \mbox{} -2 f_n(\lambda_i) f_{n+1}(\lambda_e) g_n(\lambda_e) \alpha  \mu  \beta _{n}^0 \lambda _e \lambda _i
\nonumber\\
  && \left. \mbox{} -2 f_n(\lambda_e) f_n(\lambda_i) g_{n+1}(\lambda_e) \alpha  \mu  \beta _{n}^0 \lambda _e \lambda _i-f_n(\lambda_i) g_n(\lambda_e) \alpha  \mu  \alpha _{n}^0 \lambda _e^2 \lambda _i\right) \nonumber\\
  && \mbox{} /\left(-2 f_{n+1}(\lambda_i) g_n(\lambda_e)-4 f_{n+1}(\lambda_i) g_n(\lambda_e) n+2 f_{n+1}(\lambda_i) g_n(\lambda_e) \mu +4 f_{n+1}(\lambda_i) g_n(\lambda_e) n \mu \right.\nonumber\\
  &&  \mbox{}+2 f_{n+1}(\lambda_i) g_{n+1}(\lambda_e) \lambda _e -2 f_{n+1}(\lambda_i) g_{n+1}(\lambda_e) \mu  \lambda _e+f_{n+1}(\lambda_i) g_n(\lambda_e) \lambda _e^2-f_n(\lambda_i) g_n(\lambda_e) \mu  \lambda _i\nonumber\\
  &&  \mbox{}-2 f_n(\lambda_i) g_n(\lambda_e) n \mu  \lambda _i+f_n(\lambda_i) g_{n+1}(\lambda_e) \mu  \lambda _e \lambda _i+4 f_{n+1}(\lambda_i) g_n(\lambda_e) \alpha  \mu+8 f_{n+1}(\lambda_i) g_n(\lambda_e) n \alpha  \mu\nonumber\\
  &&  \mbox{}-4 f_{n+1}(\lambda_i) g_{n+1}(\lambda_e) \alpha  \mu  \lambda _e-2 f_{n+1}(\lambda_i) g_n(\lambda_e) \alpha  \mu  \lambda _e^2-2 f_n(\lambda_i) g_n(\lambda_e) \alpha  \mu  \lambda _i\nonumber\\
  && \left. \mbox{}-4 f_n(\lambda_i) g_n(\lambda_e) n \alpha  \mu  \lambda _i+2 f_n(\lambda_i) g_{n+1}(\lambda_e) \alpha  \mu  \lambda _e \lambda _i+f_n(\lambda_i) g_n(\lambda_e) \alpha  \mu  \lambda _e^2 \lambda _i \right),
\end{eqnarray}
\begin{eqnarray}
\hat{\beta}_{n}^0&=&\left(2 f_{n+1}(\lambda_i) \alpha _{n}^0+4 f_{n+1}(\lambda_i) n \alpha _{n}^0-2 f_{n+1}(\lambda_i) \mu  \alpha _{n}^0-4 f_{n+1}(\lambda_i) n \mu  \alpha _{n}^0+2 f_n(\lambda_e)  f_{n+1}(\lambda_i) \beta _{n}^0
\right. \nonumber\\
  &&  \mbox{}
+4 f_n(\lambda_e)  f_{n+1}(\lambda_i) n \beta _{n}^0-2 f_n(\lambda_e)  f_{n+1}(\lambda_i) \mu  \beta _{n}^0-4 f_n(\lambda_e)  f_{n+1}(\lambda_i) n \mu  \beta _{n}^0\nonumber\\
  &&  \mbox{}
  +2 f_{n+1}(\lambda_e) f_{n+1}(\lambda_i) \beta _{n}^0 \lambda _e-2 f_{n+1}(\lambda_e) f_{n+1}(\lambda_i) \mu  \beta _{n}^0 \lambda _e-f_n(\lambda_e)  f_{n+1}(\lambda_i) \beta _{n}^0 \lambda _e^2
  \nonumber\\
  &&  \mbox{}
  +f_n(\lambda_i) \mu  \alpha _{n}^0 \lambda _i+2 f_n(\lambda_i) n \mu  \alpha _{n}^0 \lambda _i+f_n(\lambda_e)  f_n(\lambda_i) \mu  \beta _{n}^0 \lambda _i
  \nonumber\\
  &&  \mbox{}
  +2 f_n(\lambda_e)  f_n(\lambda_i) n \mu  \beta _{n}^0 \lambda _i+f_n(\lambda_i) f_{n+1}(\lambda_e) \mu  \beta _{n}^0 \lambda _e \lambda _i-4 f_{n+1}(\lambda_i) \alpha  \mu  \alpha _{n}^0
   \nonumber\\
  &&  \mbox{}-8 f_{n+1}(\lambda_i) n \alpha  \mu  \alpha _{n}^0-4 f_n(\lambda_e)  f_{n+1}(\lambda_i) \alpha  \mu  \beta _{n}^0-8 f_n(\lambda_e)  f_{n+1}(\lambda_i) n \alpha  \mu  \beta _{n}^0
  \nonumber\\
  &&  \mbox{}
  -4 f_{n+1}(\lambda_e) f_{n+1}(\lambda_i) \alpha  \mu  \beta _{n}^0 \lambda _e +2 f_n(\lambda_e)  f_{n+1}(\lambda_i) \alpha  \mu  \beta _{n}^0 \lambda _e^2 +2 f_n(\lambda_i) \alpha  \mu  \alpha _{n}^0 \lambda _i\nonumber\\
  &&  \mbox{}
  +4 f_n(\lambda_i) n \alpha  \mu  \alpha _{n}^0 \lambda _i +2 f_n(\lambda_e)  f_n(\lambda_i) \alpha  \mu  \beta _{n}^0 \lambda _i +4 f_n(\lambda_e)  f_n(\lambda_i) n \alpha  \mu  \beta _{n}^0 \lambda _i \nonumber\\
  && \left. \mbox{}
  +2 f_n(\lambda_i) f_{n+1}(\lambda_e) \alpha  \mu  \beta _{n}^0 \lambda _e \lambda _i -f_n(\lambda_e)  f_n(\lambda_i) \alpha  \mu  \beta _{n}^0 \lambda _e^2 \lambda _i \right)/
  \nonumber\\
  &&  \mbox{}
  \left(-2 f_{n+1}(\lambda_i) g_n(\lambda_e)-4 f_{n+1}(\lambda_i) g_n(\lambda_e) n+2 f_{n+1}(\lambda_i) g_n(\lambda_e) \mu +4 f_{n+1}(\lambda_i) g_n(\lambda_e) n \mu \right. \nonumber\\
  &&  \mbox{}+2 f_{n+1}(\lambda_i) g_{n+1}(\lambda_e) \lambda _e-2 f_{n+1}(\lambda_i) g_{n+1}(\lambda_e) \mu  \lambda _e+f_{n+1}(\lambda_i) g_n(\lambda_e) \lambda _e^2
  \nonumber\\
  &&  \mbox{}
  -f_n(\lambda_i) g_n(\lambda_e) \mu  \lambda _i-2 f_n(\lambda_i) g_n(\lambda_e) n \mu  \lambda _i+f_n(\lambda_i) g_{n+1}(\lambda_e) \mu  \lambda _e \lambda _i
  \nonumber\\
  &&  \mbox{}
  +4 f_{n+1}(\lambda_i) g_n(\lambda_e) \alpha  \mu+8 f_{n+1}(\lambda_i) g_n(\lambda_e) n \alpha  \mu  -4 f_2(\lambda_i) g_2(\lambda_e) \alpha  \mu  \lambda _e\nonumber\\
  &&  \mbox{}-2 f_{n+1}(\lambda_i) g_n(\lambda_e) \alpha  \mu  \lambda _e^2-2 f_n(\lambda_i) g_n(\lambda_e) \alpha  \mu  \lambda _i-4 f_n(\lambda_i) g_n(\lambda_e) n \alpha  \mu  \lambda _i\nonumber\\
  && \left. \mbox{}+2 f_n(\lambda_i) g_{n+1}(\lambda_e) \alpha  \mu  \lambda _e \lambda _i+f_n(\lambda_i) g_n(\lambda_e) \alpha  \mu  \lambda _e^2 \lambda _i \right),
\end{eqnarray}

\begin{eqnarray}
\bar{\alpha}_{n}^0&=&-e^{t \left(\lambda _e^2-\lambda _i^2\right)}f_n(\lambda_i)\lambda _e^2\left((g_n(\lambda_e)+2 g_n(\lambda_e) n)\alpha _{n}^0+(f_{n+1}(\lambda_e)g_n(\lambda_e)+f_n(\lambda_e) g_{n+1}(\lambda_e))\beta _{n}^0\lambda_e\right)
\nonumber\\
  &&  \mbox{}/\left(\lambda_i\left(g_n(\lambda_e)\lambda_e^2 \left(f_{n+1}(\lambda_i)+\alpha\mu \left(-2f_{n+1}(\lambda_i)+f_n(\lambda_i)\lambda_i\right)\right)\right.\right.
  \nonumber\\
  &&  \mbox{}+g_{n+1}(\lambda_e)\lambda_e \left(f_n(\lambda_i)\mu\lambda_i \left(1+2\alpha\right)-2 f_{n+1}(\lambda_i)\left(-1+\mu +2 \alpha\mu\right)\right)\nonumber\\
  && \left.\left. \mbox{}+g_n(\lambda_e)(1+2n)\left(-f_n(\lambda_i)\mu\lambda_i
\left(1+2\alpha\right)+2f_{n+1}(\lambda_i)\left(-1+\mu+2\alpha\mu\right)\right)\right)\right),
\end{eqnarray}

\begin{eqnarray}
\hat{\beta}_{n}^0&=&e^{t \left(\lambda _e^2-\lambda _i^2\right)}\lambda _e^2\left((g_n(\lambda_e)+2 g_n(\lambda_e) n)\alpha _{n}^0+(f_{n+1}(\lambda_e)g_n(\lambda_e)+f_n(\lambda_e) g_{n+1}(\lambda_e))\beta _{n}^0\lambda_e\right)\nonumber\\
 &&  \mbox{}/\left(\lambda_i\left(g_n(\lambda_e)\lambda_e^2 \left(f_{n+1}(\lambda_i)+\alpha\mu \left(-2f_{n+1}(\lambda_i)+f_n(\lambda_i)\lambda_i\right)\right)\right.\right.
  \nonumber\\
  &&  \mbox{}+g_{n+1}(\lambda_e)\lambda_e \left(f_n(\lambda_i)\mu\lambda_i \left(1+2\alpha\right)-2 f_{n+1}(\lambda_i)\left(-1+\mu +2 \alpha\mu\right)\right)\nonumber\\
  && \left.\left. \mbox{}+g_n(\lambda_e)(1+2n)\left(-f_n(\lambda_i)\mu\lambda_i
\left(1+2\alpha\right)+2f_{n+1}(\lambda_i)\left(-1+\mu+2\alpha\mu\right)\right)\right)\right),
\end{eqnarray}

\begin{eqnarray}
\hat{\gamma}_{n}^0&=&\gamma _{n}^0\left(f_{n+1}(\lambda_e)\lambda_e \left(f_n(\lambda_i)+e^{t \lambda_e^2} \alpha \mu\left(f_n(\lambda_i)(-1+n)+f_{n+1}(\lambda_i)\lambda_i\right)\mu _e\right)\right.\nonumber\\
  &&  \mbox{}+f_n(\lambda_e) \left(f_{n+1}(\lambda_i) \mu \lambda_i \left(-1+\left(-1+e^{t \lambda_e^2} n\right) \alpha  \right)\right.\nonumber\\
  && \left.\left. \mbox{}+f_n(\lambda_i)(-1+n)\left(1-\mu +\left(-1+e^{t \lambda_e^2} n\right) \alpha \mu \right)\right)\right)/\nonumber\\
  &&  \mbox{}\left(g_{n+1}(\lambda_e)\lambda_e\left(f_n(\lambda_i)+e^{t\lambda_e^2}\alpha  \mu \left(f_n(\lambda_i) (-1+n)+f_{n+1}(\lambda_i) \lambda_i\right) \mu _e\right)\right.\nonumber\\
  &&  \mbox{}-g_n(\lambda_e) \left(f_{n+1}(\lambda_i) \mu  \lambda _i \left(-1+\left(-1+e^{t \lambda_e^2} n\right) \alpha  \right)\right.\nonumber\\
  &&  \left.\left. \mbox{}+f_n(\lambda_i) (-1+n) \left(1-\mu +\left(-1+e^{t \lambda_e^2} n\right) \alpha  \mu  \right)\right)\right),
\end{eqnarray}
\begin{eqnarray}
\hat{\gamma}_{n}^0&=&\left(e^{t \left(\lambda _e^2-\lambda _i^2\right)}\gamma _{n}^0(f_{n+1}(\lambda_e) g_n(\lambda_e)+f_n(\lambda_e)g_{n+1}(\lambda_e)) \lambda _e \left(-1+\left(-1+e^{t \lambda _e^2}\right) \alpha\right)\right)\nonumber\\
  &&  \mbox{}/\left(-g_{n+1}(\lambda_e)\lambda _e \left(f_n(\lambda_i)+e^{t \lambda _e^2} \alpha  \mu  \left(f_n(\lambda_i)(-1+n)+f_{n+1}(\lambda_i)\lambda _i\right) \mu _e\right)\right.\nonumber\\
  &&  \mbox{}+g_n(\lambda_e) \left(f_{n+1}(\lambda_i)\mu  \lambda _i \left(-1+\left(-1+e^{t \lambda _e^2} n\right) \alpha\right)\right.\nonumber\\
  && \left.\left. \mbox{}+f_n(\lambda_i) (-1+n) \left(1-\mu +\left(-1+e^{t \lambda _e^2} n\right) \alpha  \mu\right)\right)\right).
\end{eqnarray}
\section{Symbols}\label{appB}
The constants given in (\ref{eq:29}) are given as follows:
\[
X=\lambda_e^3\{\lambda_i
f_1(\lambda_i)-2f_2(\lambda_i)\}g_2(\lambda_e),
\]
\[
Y=\lambda_e^3\{\lambda_i
g_1(\lambda_e)+2g_2(\lambda_e)\}f_2(\lambda_i),
\]
\[
P=-\mu\lambda_e^3\{-\lambda_if_1(\lambda_i)+2f_2(\lambda_i)\}\{2g_2(\lambda_e)+
g_1(\lambda_e)\},
\]
\[
G=\mu\{2g_2(\lambda_e)\lambda_e-6g_1(\lambda_e)+\lambda_e^2g_1(\lambda_e)\}
\{\lambda_if_1(\lambda_i)-2f_2(\lambda_i)\},
\]
\[
Z=\{\lambda_if_1(\lambda_i)-2f_2(\lambda_i)\}\{\lambda_e
g_2(\lambda_e)-3g_1(\lambda_e)\},
\]
\[
W=\{2\lambda_e
g_2(\lambda_e)-(6-\lambda_e^2)g_1(\lambda_e)\}f_2(\lambda_i),
\]
\[
S=3\{f_2(\lambda_e)g_1(\lambda_e)\}\{\lambda_i
f_1(\lambda_i)-2f_2(\lambda_i)\},
\]
\[
T=6f_2(\lambda_i)\{f_2(\lambda_e)g_1(\lambda_e)+f_1(\lambda_e)g_2(\lambda_e)\},
\]
\[
Q=-3\mu\{f_2(\lambda_e)g_1(\lambda_e)+f_1(\lambda_e)g_2(\lambda_e)\}
\{4f_2(\lambda_i)-2\lambda_if_2(\lambda_i)\},
\]
\[
U=S-\frac{X}{\lambda_e^2}; V=T-\frac{Y}{\lambda_e^2};  H=Q-\frac{P}{\lambda_e^2}.
\]

\section{The unknown coefficients (for first order problem)}\label{appC}
The unknown coefficients given in (\ref{eq:21f}) to (\ref{eq:24f}) are given by
\begin{eqnarray}
\bar{\alpha} _{n}^1&=&\left(e^{t \lambda _e^2-t \lambda _i^2} f_n(\lambda_i) {Ma} \left(g_n(\lambda_e)+2 g_n(\lambda_e) n-g_{n+1}(\lambda_e) \lambda _e+2 g_n(\lambda_e) \alpha \right.\right. \nonumber\\
  && \left.\left. \mbox{}+4 g_n(\lambda_e) n \alpha -2 g_{n+1}(\lambda_e) \alpha  \lambda _e -g_n(\lambda_e) \alpha  \lambda _e^2 \right)\right)\nonumber\\
  &&  \mbox{}/\left(\lambda _i \left(2 f_{n+1}(\lambda_i) g_n(\lambda_e)+4 f_{n+1}(\lambda_i) g_n(\lambda_e) n-2 f_{n+1}(\lambda_i) g_n(\lambda_e) \mu -4 f_{n+1}(\lambda_i) g_n(\lambda_e) n \mu\right.\right.
  \nonumber\\
  &&  \mbox{}
   -2 f_{n+1}(\lambda_i) g_{n+1}(\lambda_e) \lambda _e+2 f_{n+1}(\lambda_i) g_{n+1}(\lambda_e) \mu  \lambda _e-f_{n+1}(\lambda_i) g_n(\lambda_e) \lambda _e^2+f_n(\lambda_i) g_n(\lambda_e) \mu  \lambda _i
   \nonumber\\
  &&  \mbox{}+2 f_n(\lambda_i) g_n(\lambda_e) n \mu  \lambda _i-f_n(\lambda_i) g_{n+1}(\lambda_e) \mu  \lambda _e \lambda _i-4 f_{n+1}(\lambda_i) g_n(\lambda_e) \alpha  \mu
  \nonumber\\
  &&  \mbox{}-8 f_{n+1}(\lambda_i) g_n(\lambda_e) n \alpha  \mu +4 f_{n+1}(\lambda_i) g_{n+1}(\lambda_e) \alpha  \mu  \lambda _e +2 f_{n+1}(\lambda_i) g_n(\lambda_e) \alpha  \mu  \lambda _e^2 \nonumber\\
  &&  \mbox{}+2 f_n(\lambda_i) g_n(\lambda_e) \alpha  \mu  \lambda _i +4 f_n(\lambda_i) g_n(\lambda_e) n \alpha  \mu  \lambda _i \nonumber\\
  && \left.\left. \mbox{}-2 f_n(\lambda_i) g_{n+1}(\lambda_e) \alpha  \mu  \lambda _e \lambda _i -f_n(\lambda_i) g_n(\lambda_e) \alpha  \mu  \lambda _e^2 \lambda _i \right)\right),
\end{eqnarray}
\begin{eqnarray}
\bar{\beta} _{n}^1&=&-\left(e^{t \lambda _e^2-t \lambda _i^2} {Ma} \left(g_n(\lambda_e)+2 g_n(\lambda_e) n-g_{n+1}(\lambda_e) \lambda _e+2 g_n(\lambda_e) \alpha\right.\right. \nonumber\\
  && \left.\left. \mbox{}+4 g_n(\lambda_e) n \alpha -2 g_{n+1}(\lambda_e) \alpha  \lambda _e -g_n(\lambda_e) \alpha  \lambda _e^2 \right)\right)\nonumber\\
  &&  \mbox{}/\left(\lambda _i \left(2 f_{n+1}(\lambda_i) g_n(\lambda_e)+4 f_{n+1}(\lambda_i) g_n(\lambda_e) n-2 f_{n+1}(\lambda_i) g_n(\lambda_e) \mu -4 f_{n+1}(\lambda_i) g_n(\lambda_e) n \mu\right.\right.
  \nonumber\\
  &&  \mbox{}
   -2 f_{n+1}(\lambda_i) g_{n+1}(\lambda_e) \lambda _e+2 f_{n+1}(\lambda_i) g_{n+1}(\lambda_e) \mu  \lambda _e-f_{n+1}(\lambda_i) g_n(\lambda_e) \lambda _e^2+f_n(\lambda_i) g_n(\lambda_e) \mu  \lambda _i
   \nonumber\\
  &&  \mbox{}+2 f_n(\lambda_i) g_n(\lambda_e) n \mu  \lambda _i-f_n(\lambda_i) g_{n+1}(\lambda_e) \mu  \lambda _e \lambda _i-4 f_{n+1}(\lambda_i) g_n(\lambda_e) \alpha  \mu
  \nonumber\\
  &&  \mbox{}-8 f_{n+1}(\lambda_i) g_n(\lambda_e) n \alpha  \mu +4 f_{n+1}(\lambda_i) g_{n+1}(\lambda_e) \alpha  \mu  \lambda _e +2 f_{n+1}(\lambda_i) g_n(\lambda_e) \alpha  \mu  \lambda _e^2 \nonumber\\
  &&  \mbox{}+2 f_n(\lambda_i) g_n(\lambda_e) \alpha  \mu  \lambda _i +4 f_n(\lambda_i) g_n(\lambda_e) n \alpha  \mu  \lambda _i \nonumber\\
  && \left.\left. \mbox{}-2 f_n(\lambda_i) g_{n+1}(\lambda_e) \alpha  \mu  \lambda _e \lambda _i -f_n(\lambda_i) g_n(\lambda_e) \alpha  \mu  \lambda _e^2 \lambda _i \right)\right),
\end{eqnarray}
\begin{eqnarray}
\bar{\gamma} _{n}^1=0,
\end{eqnarray}
\begin{eqnarray}
\alpha _{n}^1=0,
\end{eqnarray}
\begin{eqnarray}
\hat{\alpha} _{n}^1&=&\left(f_{n+1}(\lambda_i) g_n(\lambda_e) {Ma}\right)\nonumber\\
  &&  \mbox{}
  /\left( 2 f_{n+1}(\lambda_i) g_n(\lambda_e)+4 f_{n+1}(\lambda_i) g_n(\lambda_e) n-2 f_{n+1}(\lambda_i) g_n(\lambda_e) \mu -4 f_{n+1}(\lambda_i) g_n(\lambda_e) n \mu\right.
  \nonumber\\
  &&  \mbox{}
   -2 f_{n+1}(\lambda_i) g_{n+1}(\lambda_e) \lambda _e+2 f_{n+1}(\lambda_i) g_{n+1}(\lambda_e) \mu  \lambda _e-f_{n+1}(\lambda_i) g_n(\lambda_e) \lambda _e^2+f_n(\lambda_i) g_n(\lambda_e) \mu  \lambda _i
   \nonumber\\
  &&  \mbox{}+2 f_n(\lambda_i) g_n(\lambda_e) n \mu  \lambda _i-f_n(\lambda_i) g_{n+1}(\lambda_e) \mu  \lambda _e \lambda _i-4 f_{n+1}(\lambda_i) g_n(\lambda_e) \alpha  \mu
  \nonumber\\
  &&  \mbox{}-8 f_{n+1}(\lambda_i) g_n(\lambda_e) n \alpha  \mu +4 f_{n+1}(\lambda_i) g_{n+1}(\lambda_e) \alpha  \mu  \lambda _e +2 f_{n+1}(\lambda_i) g_n(\lambda_e) \alpha  \mu  \lambda _e^2 \nonumber\\
  &&  \mbox{}+2 f_n(\lambda_i) g_n(\lambda_e) \alpha  \mu  \lambda _i +4 f_n(\lambda_i) g_n(\lambda_e) n \alpha  \mu  \lambda _i \nonumber\\
  && \left. \mbox{}-2 f_n(\lambda_i) g_{n+1}(\lambda_e) \alpha  \mu  \lambda _e \lambda _i -f_n(\lambda_i) g_n(\lambda_e) \alpha  \mu  \lambda _e^2 \lambda _i \right),
\end{eqnarray}
\begin{eqnarray}
{\beta} _{n}^1=0,
\end{eqnarray}
\begin{eqnarray}
\hat{\beta} _{n}^1&=&-\left(f_{n+1}(\lambda_i) {Ma}\right)\nonumber\\
  &&  \mbox{}/\left( 2 f_{n+1}(\lambda_i) g_n(\lambda_e)+4 f_{n+1}(\lambda_i) g_n(\lambda_e) n-2 f_{n+1}(\lambda_i) g_n(\lambda_e) \mu -4 f_{n+1}(\lambda_i) g_n(\lambda_e) n \mu\right.
  \nonumber\\
  &&  \mbox{}
   -2 f_{n+1}(\lambda_i) g_{n+1}(\lambda_e) \lambda _e+2 f_{n+1}(\lambda_i) g_{n+1}(\lambda_e) \mu  \lambda _e-f_{n+1}(\lambda_i) g_n(\lambda_e) \lambda _e^2+f_n(\lambda_i) g_n(\lambda_e) \mu  \lambda _i
   \nonumber\\
  &&  \mbox{}+2 f_n(\lambda_i) g_n(\lambda_e) n \mu  \lambda _i-f_n(\lambda_i) g_{n+1}(\lambda_e) \mu  \lambda _e \lambda _i-4 f_{n+1}(\lambda_i) g_n(\lambda_e) \alpha  \mu
  \nonumber\\
  &&  \mbox{}-8 f_{n+1}(\lambda_i) g_n(\lambda_e) n \alpha  \mu +4 f_{n+1}(\lambda_i) g_{n+1}(\lambda_e) \alpha  \mu  \lambda _e +2 f_{n+1}(\lambda_i) g_n(\lambda_e) \alpha  \mu  \lambda _e^2 \nonumber\\
  &&  \mbox{}+2 f_n(\lambda_i) g_n(\lambda_e) \alpha  \mu  \lambda _i +4 f_n(\lambda_i) g_n(\lambda_e) n \alpha  \mu  \lambda _i \nonumber\\
  && \left. \mbox{}-2 f_n(\lambda_i) g_{n+1}(\lambda_e) \alpha  \mu  \lambda _e \lambda _i -f_n(\lambda_i) g_n(\lambda_e) \alpha  \mu  \lambda _e^2 \lambda _i \right),
\end{eqnarray}
\begin{eqnarray}
{\gamma} _{n}^1=0,
\end{eqnarray}
\begin{eqnarray}
\hat{\gamma} _{n}^1=0.
\end{eqnarray}

\section{Coefficients Couette flow}\label{appD}

\begin{eqnarray}
E_{11}^1=\frac{2FL}{(2+\lambda_e^2Pr_s)}\frac{\left(3 g_1(\lambda_e) \lambda _e^2 \left(f_2(\lambda_i)+\alpha  \mu  \left(-2 f_2(\lambda_i)+f_1(\lambda_i) \lambda _i\right) \right)\right)}{\delta_1},
 \label{eq:couette3}
\end{eqnarray}
\begin{eqnarray}
F_{22}^1=\frac{5 F g_2(\lambda_e) \lambda _e^2 \left(f_3(\lambda_i) (2 \alpha  \mu -1)-\alpha  f_2(\lambda_i) \mu  \lambda _i\right)}{6 \left(Pr_s  \lambda _e^2+6\right)\delta_2}
\end{eqnarray}

where
\begin{eqnarray}
\delta_2&=&g_2(\lambda_e) \lambda _e^2 \left(f_3(\lambda_i) (2 \alpha  \mu -1)-\alpha  f_2(\lambda_i) \mu  \lambda _i\right)
\nonumber\\  && \mbox{}
+g_3(\lambda_e) \lambda _e \left(2 f_3(\lambda_i) (2 \alpha  \mu +\mu -1)-(2 \alpha +1) f_2(\lambda_i) \mu  \lambda _i\right) \nonumber\\  && \mbox{}
+5 g_2(\lambda_e) \left((2 \alpha +1) f_2(\lambda_i) \mu  \lambda _i-2 f_3(\lambda_i) (2 \alpha  \mu +\mu -1)\right),
\end{eqnarray}
and
\begin{eqnarray}
E_{11}^2& = &\frac{4 Ma
E_{11}^1}{(2+\lambda_e^2Pr_s)}\frac{f_2(\lambda_i)\left(-3
g_1(\lambda_e)+g_2(\lambda_e)\lambda_e\right)}{\delta_1},
 \label{eq:couette4}
\end{eqnarray}

\begin{eqnarray}
F_{11}^2& = &\frac{\lambda _e^2 e^{\lambda_e^2 t}}{5 \left(2+Pr_s
\lambda _e^2\right)}
 \nonumber\\  &&
\mbox{}
 \left(\frac{18 F_{22}^1 g_1(\lambda_e) (F L) \left(-\alpha f_1(\lambda_i) \mu  \lambda _i+f_2(\lambda_i) (2 \alpha  \mu -1)\right)}{\delta_3}\right.
\nonumber\\  && \left.\mbox{} +\frac{5 E_{11}^1 F g_2(\lambda_e)
\left(-\alpha  f_2(\lambda_i) \mu  \lambda _i+f_3(\lambda_i) (2
\alpha  \mu -1)\right)}{\delta_2}\right),
\end{eqnarray}
where

\begin{eqnarray}
\delta_3&=&g_1(\lambda_e) \lambda _e^2 \left(\alpha  f_1(\lambda_i) \mu  \lambda _i-2 \alpha  f_2(\lambda_i) \mu +f_2(\lambda_i)\right)\nonumber\\  && \mbox{}
+g_2(\lambda_e) \lambda _e \left((2 \alpha +1) f_1(\lambda_i) \mu  \lambda _i-2 f_2(\lambda_i) (2 \alpha  \mu +\mu -1)\right)\nonumber\\  && \mbox{}
+3 g_1(\lambda_e) \left(2 f_2(\lambda_i) (2 \alpha  \mu +\mu -1)-(2 \alpha +1) f_1(\lambda_i) \mu  \lambda _i\right)
\end{eqnarray}
\begin{eqnarray}
F_{22}^2& = &\frac{6 F_{22}^1 f_3(\lambda_i) Ma \left(-5
g_2(\lambda_e)+g_3(\lambda_e) \lambda _e\right)}{\left(\Pr
\lambda _e^2+6\right)\delta_2},
 \label{eq:couette6}
\end{eqnarray}
\begin{eqnarray}
F_{31}^2& = &\frac{2\lambda _e^2 e^{\lambda_e^2 t}}{45
\left(12+Pr_s  \lambda _e^2\right)}
 \nonumber\\  &&
\mbox{}
 \left(-\frac{27 F_{22}^1 g_1(\lambda_e) (F L) \left(-\alpha f_1(\lambda_i) \mu  \lambda _i+f_2(\lambda_i) (2 \alpha  \mu -1)\right)}{\delta_3}\right.
\nonumber\\  && \left.\mbox{} -\frac{5 E_{11}^1 F g_2(\lambda_e)
\left(-\alpha  f_2(\lambda_i) \mu  \lambda _i+f_3(\lambda_i) (2
\alpha  \mu -1)\right)}{\delta_2}\right),
 \label{eq:couette7}
\end{eqnarray}
\begin{eqnarray}
F_{33}^2& = &\frac{-\lambda _e^2 e^{\lambda_e^2 t}}{45
\left(12+Pr_s  \lambda _e^2\right)}
 \nonumber\\  &&
\mbox{}
 \left(-\frac{27 F_{22}^1 g_1(\lambda_e) (F L) \left(-\alpha f_1(\lambda_i) \mu  \lambda _i+f_2(\lambda_i) (2 \alpha  \mu -1)\right)}{\delta_3}\right.
\nonumber\\  && \left.\mbox{} -\frac{5 E_{11}^1 F g_2(\lambda_e)
\left(-\alpha  f_2(\lambda_i) \mu  \lambda _i+f_3(\lambda_i) (2
\alpha  \mu -1)\right)}{\delta_2}\right),
 \label{eq:couette8}
\end{eqnarray}

\begin{eqnarray}
E_{20}^2& = &\frac{\lambda _e^2 e^{\lambda_e^2 t}}{7 \left(6+Pr_s
\lambda _e^2\right)}
 \nonumber\\  &&
\mbox{}
 \left(\frac{-20 F_{22}^1 F g_2(\lambda_e) \left(-\alpha  f_2(\lambda_i) \mu  \lambda _i+f_3(\lambda_i) (2 \alpha  \mu -1)\right)}{\delta_2}\right.
\nonumber\\  && \left.\mbox{} +\frac{21 E_{11}^1 (F L)g_1(\lambda_e)
 \left(-\alpha f_1(\lambda_i) \mu  \lambda _i+f_2(\lambda_i) (2
\alpha  \mu -1)\right)}{\delta_3}\right),
 \label{eq:couette9}
\end{eqnarray}

\begin{eqnarray}
E_{22}^2& = &\frac{-9\lambda _e^2 e^{\lambda_e^2 t}}{\left(6+Pr_s
\lambda _e^2\right)}
 \nonumber\\  &&
\mbox{}
 \left(\frac{E_{11}^1 (F L)g_1(\lambda_e)  \left(-\alpha f_1(\lambda_i) \mu  \lambda _i+f_2(\lambda_i) (2 \alpha  \mu -1)\right)}{\delta_3}\right),
 \label{eq:couette10}
\end{eqnarray}

\begin{eqnarray}
E^2_{40}& = &\frac{\lambda _e^2 e^{\lambda_e^2 t}}{7 \left(20+Pr_s
\lambda _e^2\right)}
 \nonumber\\  &&
\mbox{}
 \left(\frac{20 F_{22}^1 F g_2(\lambda_e) \left(-\alpha  f_2(\lambda_i) \mu  \lambda _i+f_3(\lambda_i) (2 \alpha  \mu -1)\right)}{\delta_2}\right),
 \label{eq:couette11}
\end{eqnarray}

\begin{eqnarray}
E^2_{44}& = &\frac{-\lambda _e^2 e^{\lambda_e^2 t}}{84
\left(20+Pr_s  \lambda _e^2\right)}
 \nonumber\\  &&
\mbox{}
 \left(\frac{5 F_{22}^1 F g_2(\lambda_e) \left(-\alpha  f_2(\lambda_i) \mu  \lambda _i+f_3(\lambda_i) (2 \alpha  \mu -1)\right)}{\delta_2}\right),
 \label{eq:couette11}
\end{eqnarray}

\bibliographystyle{IEEEtran}
\bibliography{A}

\end{document}